\documentclass[10pt,superscriptaddress,english,showpacs,longbibliography,nofootinbib,floatfix]{revtex4-1}
\pdfoutput=1
\usepackage{amsmath}
\usepackage{amssymb}
\usepackage{amsthm}
\usepackage{amsfonts}
\usepackage{listings}
\usepackage{longtable}
\usepackage{enumerate}
\usepackage{latexsym}
\usepackage{color}
\usepackage{multirow}
\usepackage[figuresright]{rotating}
\usepackage[table]{xcolor}
\usepackage{setspace} 
\usepackage{blindtext}
\usepackage{dcolumn}
\usepackage{braket}

\usepackage{array,etoolbox}
\preto\tabular{\setcounter{magicrownumbers}{0}}
\newcounter{magicrownumbers}

\usepackage{bm}
\usepackage{hyperref}
\hypersetup{
 pdfnewwindow=true, colorlinks=true,
 linkcolor=blue, anchorcolor=blue,
 citecolor=blue, filecolor=blue,
 menucolor=blue, urlcolor=blue}
\definecolor{red}{rgb}{1,0,0}
\definecolor{blue}{rgb}{0,0,1}
\definecolor{black}{rgb}{0,0,0}

\newcommand{\sq}{$^2$}
\newcommand{\vsim}{[V$^2$]}
\newcommand{\msim}{[M$^2$]}
\newcommand{\oal}{\overline{\alpha}}
\newcommand{\st}{$^3$}
\newcommand{\Gso}{G$_{\textrm{\footnotesize{SO}}}$}
\newcommand{\Gss}{G$_{\textrm{\footnotesize{SS}}}$}
\newcommand{\Gnt}{G$_{\textrm{\footnotesize{NT}}}$}
\newcommand{\Gst}{G$_{\textrm{\footnotesize{ST}}}$}
\newcommand{\Gp}{G$_{\textrm{\footnotesize{P}}}$}
\newcommand{\ssgcol}{$^{\infty_{\bf n}m}1$}
\newcommand{\ssgcop}{$^{m_{\bf n}}1$}
\newcommand{\ssgcolt}{^{\infty_{\bf n}m}1}
\newcommand{\ssgcopt}{^{m_{\bf n}}1}
\newcommand{\Pso}{P$_{\textrm{\footnotesize{SO}}}$}
\newcommand{\Ps}{P$_{\textrm{\footnotesize{S}}}$}
\newcommand{\Pm}{P$_{\textrm{\footnotesize{M}}}$}
\newcommand{\Pnt}{P$_{\textrm{\footnotesize{NT}}}$}
\newcommand{\Psoin}{P$_{\textrm{\footnotesize{SOintr}}}$}
\newcommand{\Psog}{P$_{\textrm{\footnotesize{SOG}}}$}
\newcommand{\mpgeff}{MPG$_\textrm{\footnotesize{eff}}$}

 \usepackage{appendix}

\begin{document}

\title{Crystal tensor properties of magnetic materials with and without spin-orbit coupling. Application of spin point groups as approximate symmetries.}

\author{Jesus Etxebarria}
	\email{j.etxeba@ehu.eus}
\affiliation{Department of Physics, Faculty of Science and Technology, UPV/EHU, Bilbao, Spain}

\author{J. Manuel Perez-Mato}
\email{jm.perezmato@gmail.com}
\affiliation{Faculty of Science and Technology, UPV/EHU, Bilbao, Spain}

	\author{Emre S. Tasci}
\affiliation{Department of Physics Engineering, Hacettepe University, 06800 Ankara, Turkey}
\author{Luis Elcoro}
\affiliation{Department of Physics, Faculty of Science and Technology, UPV/EHU, Bilbao, Spain}

\begin{abstract}
Spin space groups, formed by operations where the rotation of the spins is independent of the accompanying operation acting on the crystal structure, are appropriate groups to describe the symmetry of magnetic structures with null spin-orbit coupling. Their corresponding spin point groups are the symmetry groups to be considered for deriving the symmetry constraints on the form of the crystal tensor properties of such idealized structures. These groups can also be taken as approximate symmetries (with some restrictions) of real magnetic structures, where spin-orbit and magnetic anisotropy are however present.  Here we formalize the invariance transformation properties that must satisfy the most important crystal tensors under a spin point group. This is done using modified Jahn symbols, which generalize those applicable to ordinary magnetic point groups [Gallego et al., Acta Cryst. (2019) A{\bf 75}, 438-447]. The analysis includes not only equilibrium tensors, but also transport, optical and non-linear optical susceptibility tensors. The constraints imposed by spin collinearity and coplanarity within the spin group formalism on a series of representative tensors are discussed and compiled. As illustrative examples, the defined tensor invariance equations have been applied to some known magnetic structures, showing the differences of the symmetry-adapted form of some relevant tensors, when considered under the constraints of its spin point group or its magnetic point group. This comparison, with the spin point group implying additional constraints in the tensor form, can allow one to distinguish those magnetic-related properties that can be solely attributed to spin-orbit coupling from those that are expected even when spin-orbit coupling is negligible.
\end{abstract}

\maketitle

\section{Introduction}
\label{s-introduction}	
	Although the theory of spin space groups (SpSGs)\footnote{We do not follow the acronym used so far, as SSG have been used since many years for SuperSpace Groups within the field of crystallographic symmetry and Spin SuperSpace Groups is a possible concept to be used in the future} was proposed and developed more than fifty years ago \cite{Brinkman1966,Litvin1974,Litvin1977}, it is only recently that these groups have become the object of much interest and have been intensively applied in the frame of electronic band studies of magnetic materials. As symmetry groups associated with negligible spin-orbit coupling (SOC), the SpSG of a magnetic structure is in general a supergroup of its magnetic space group (MSG), and as consequence the SpSG may dictate symmetry constraints on the properties of the material, additional to those resulting from its MSG. In the framework of electronic bands, more symmetry constraints in general imply more band degeneracies. Thus, the application of SpSGs has been used to identify spin band splittings, which are present not only considering the MSG of the structure, but also its SpSG, and therefore they may be considered quite robust and especially important as SOC-free or non-relativistic effects \cite{Liu2022}. For example, the so-called altermagnets, which refer to collinear antiferromagnets with spin splitting in the SOC-free limit \cite{Yuan2020,Smejkal2022a,Smejkal2022b,Mazin2022}, can be described in terms of their SpSGs. The same can be said for other forms of unconventional magnetism in materials with non-collinear magnetism \cite{Yuan2021,Hellenes2024}. It is in this context that three independent groups have very recently enumerated and classified the SpSGs, and considered in detail their application in the symmetry analysis of electronic bands of magnetic materials \cite{Chen2024,Jiang2024,Xiao2024}.
	
	In general, the comparison of the SpSG and the MSG of a magnetic structure could be used to distinguish and resolve features and properties that are only SOC effects, and therefore they are expected in general to be quite weak or even negligible. In practice, for real materials, this approach may partially fail if the observed spin arrangement includes features due to SOC effects. Notwithstanding this problem, the relevance of SpSGs in the study of tensor properties of magnetic materials, establishing a general rigorous formalism, is still a field to be explored in detail. Some recent contributions along these lines have already been made \cite{Watanabe2024,Zhu2024}. This work is a further step in this direction.
	
	In \citet{Gallego2019} a comprehensive analysis of the symmetry-adapted form of all kinds of crystal tensor properties in non-magnetic and magnetic materials was performed, considering their relevant symmetry groups, namely crystallographic point groups and crystallographic magnetic point groups, respectively. Here, following a similar approach, we analyze the symmetry-adapted forms of crystal tensors under the spin point group (SpPG) associated with the SpSG of a structure, and compare them with those to be expected from its actual magnetic point group (MPG).
	
	The article is organized in the following form: after a recapitulation of the physical meaning and mathematical structure of the SpSGs and their corresponding SpPGs, their relation with ordinary MPGs is discussed in detail. We then formalize the symmetry conditions to be satisfied by crystal tensors in magnetic crystals under a given SpPG. For this purpose, the Jahn symbols \cite{Jahn1949}, describing the transformation properties of each tensor for the symmetry operations, are here generalized to take into account the particular features of SpPG operations. Using this generalization, we establish the corresponding generalized Jahn symbol for all kinds of tensors, including equilibrium, transport and optical properties. This formalism is then applied to a series of examples of experimental magnetic structures, for which the symmetry-adapted form of various tensors under the SpPG of the structure is determined, and compared with the less stringent constraints under its MPG, where possible SOC effects are necessarily taken into account. Very different types of SpPG-MPG relations can be realized in a magnetic structure, and the examples presented here try to cover the most representative ones. Finally, in section \ref{glossary} we have included a glossary of some important groups used in the paper and their notation.

	\section{Spin Space Groups and Spin Point Groups}
	\label{s-spin-groups}
	\subsection{Spin Space Groups as the Symmetry Groups of SOC-Free Magnetic Structures}
	\label{s-spin-space-groups}
	
	A well-defined symmetry group of a physical system must be constituted by operations which, apart from keeping the system indistinguishable, constitute a subgroup of the group of transformations that keep the energy of the system invariant. This ensures that the constraints on the system implied by these operations are \textit{stable}, in the sense that they are maintained if, for instance, in the case of a thermodynamic system, temperature or pressure are varied (excluding a symmetry-breaking phase transition taking place); or in the case of a system ground state, the symmetry constraints are maintained if the Hamiltonian parameters are continuously varied. As a consequence, a symmetry group defined under this condition can be assigned to a whole thermodynamic phase, or to the ground state for some continuous range of the Hamiltonian parameters. This is why in non-magnetic commensurate crystal structures the operations of the space groups, which describe their symmetry, are formed by combinations of rotations, translations, and space inversion, which all keep the energy invariant. Hereafter, we shall call this type of operations \textit{space operations}, and they will be generally represented by the symbol $\{R | \mathbf{t}\}$, where $R$ represents a proper or improper rotation of the system, including the limiting cases of $R$ being the identity $1$, or the space inversion $\overline{1}$, while $\mathbf{t}$ represents a space translation of the system.
	
	In the case of incommensurate modulated crystal structures, global phase shift(s) of the incommensurate modulation(s) also keep the energy invariant, and therefore, the so-called \emph{superspace groups} describing the symmetry of these systems are constructed by adding these extra energy-invariant transformations, when the combined symmetry operations that keep the system indistinguishable are defined 
	\cite{Janssen2004}. For instance, a generic operation of a $(3+1)$D superspace group with a single independent incommensurate wave vector can be expressed as $\{R | \mathbf{t},\tau\}$, indicating that the space operation $\{R | \mathbf{t}\}$ is followed by a global shift $\tau$ of the incommensurate modulation in the structure \cite{PerezMato1984}.
	
	In the same way, in the case of commensurate magnetic structures, MSGs are constructed by adding the time-reversal operation, which reverses both spins and momenta, when defining the operations of the group \cite{Litvin2016,Campbell2024}. The time-reversal operation indeed keeps the energy invariant, and is in fact a trivial symmetry operation always present (and therefore not explicitly considered) in all non-magnetic or magnetically disordered structures, while in magnetic structures it may only be present if combined with some space operation different from the trivial identity. Thus, a generic operation of an MSG can be expressed as $\{R,\theta|\mathbf{t}\}$, with $\theta$ being -1 if time-reversal is included, and +1 otherwise. It is important in the context of the present work to stress that the space operation $\{R,\theta | \mathbf{t}\}$ of an MSG, necessarily operates on the system as a whole, i.e., it includes also a transformation of its atomic spins or its spin density, as the spin orientation and the crystal structure are in general energy-coupled through the spin-orbit coupling (SOC). Thus, an energy-invariant operation $\{R, \theta | \mathbf{t}\}$ transforms not only the crystal structure, given for instance by a scalar density $\rho(\mathbf{r})$, with the space operation $\{R | \mathbf{t}\}$:
	\begin{equation}
		\label{e-density}
		\rho'(\mathbf{r}) = \rho\left(\{R | \mathbf{t}\}^{-1} \mathbf{r}\right),
	\end{equation}
	but it also transforms the magnetic moment density $\mathbf{M}(\mathbf{r})$ of the system into a new one $\mathbf{M}'(\mathbf{r})$ that satisfies:
	\begin{equation}
		\label{e-Mmoment}
		\mathbf{M}'(\mathbf{r}) = \theta \det(R) R \cdot \mathbf{M}\left(\{R | \mathbf{t}\}^{-1} \mathbf{r}\right),
	\end{equation}
	where $\det(R)$ is the determinant of the matrix $R$. Thus, in equation (\ref{e-Mmoment}), both the axial-vector character of the magnetic moment and the inclusion or not of time-reversal in the operation are taken into account. If after applying the operation the transformed functions coincide with the original ones, so that $\rho'(\mathbf{r}) = \rho(\mathbf{r})$ and $\mathbf{M}'(\mathbf{r}) = \mathbf{M}(\mathbf{r})$, then the operation $\{R, \theta | \mathbf{t}\}$ belongs to the MSG of the structure.
	
	MSGs are therefore the appropriate groups that can describe the symmetry of a commensurate magnetic structure, i.e., the set of symmetry constraints that are expected to be satisfied by the structure within the whole range of a thermodynamic phase, or in the case of a ground state, to be satisfied within a continuous range of the Hamiltonian parameters. However, if the SOC in the structure can be considered negligible, then any arbitrary global rotation $R_S$ of the spin arrangement, with full independence of the crystal orientation, is also energy-invariant. Here, however, we must explicitly separate the usually small orbital contribution $\mathbf{M}_{orb}(\mathbf{r})$ to the magnetization density $\mathbf{M}(\mathbf{r})$, from the contribution of the actual spins $\mathbf{M}_s(\mathbf{r})$, because these additional energy-free spin rotations to be included refer only to $\mathbf{M}_s(\mathbf{r})$, while the orbital contribution $\mathbf{M}_{orb}(\mathbf{r})$ remains locked to the space operations. Hence, we can express these additional energy-invariant transformations of $\mathbf{M}_s(\mathbf{r})$ as:
	\begin{equation}
		\label{e-Trmoment}
		\mathbf{M}_s'(\mathbf{r}) = R_s \cdot \mathbf{M}_s(\mathbf{r}),
	\end{equation}
	where $R_s$ is any 3D proper rotation. This extension in SOC-free structures of the set of energy-invariant transformations implies that their symmetry can be described by the spin space groups (SpSGs) \cite{Brinkman1966,Litvin1974}, where operations of the type considered in MSGs can also be combined with spin rotation operations of the type indicated in equation (\ref{e-Trmoment}). Thus, a generic operation of an SpSG could be expressed as $\{R_S || \{R, \theta | \mathbf{t}\}\}$, indicating the combination of an MSG-type operation $\{R,\theta | \mathbf{t}\}$ with an additional proper rotation $R_S$ of the spins. As in SOC-free structures spins are uncoupled with the crystal structure, $R_S$ in the operation above can be defined in such a way that it includes the necessary rotation to be applied to the spins, while the space operation $\{R, \theta | \mathbf{t}\}$, in contrast with its interpretation in an MSG, does not act on the spins, but applies only to the magnetic moments of orbital origin.
	
	Hence, if $\{R_S || \{R, \theta | \mathbf{t}\}\}$ is an operation of the SpSG of a magnetic structure, it implies that the following equations are fulfilled:
	\begin{eqnarray}
		\label{e-SpSGtransf-space}
		\rho(\mathbf{r}) &=& \rho\left(\{R | \mathbf{t}\}^{-1} \mathbf{r}\right), \\
		\label{e-SpSGtransf-orbital}
		\mathbf{M}_{orb}(\mathbf{r}) &=& \theta \det(R) R \cdot \mathbf{M}_{orb}\left(\{R | \mathbf{t}\}^{-1} \mathbf{r}\right), \\
		\label{e-SpSGtransf-spin}
		\mathbf{M}_s(\mathbf{r}) &=& \theta R_S \cdot \mathbf{M}_s\left(\{R | \mathbf{t}\}^{-1} \mathbf{r}\right).
	\end{eqnarray}
	
	Thus, the rotation applied to the spins is fully unlocked from the space operation and can be an improper one, $-R_S$, or a proper one, $R_S$, depending on whether the operation includes time-reversal or not. In contrast, the atomic magnetic moments of orbital origin are locked to the crystal and are transformed in the usual form of an MSG operation.
	
	For convenience, following the usual convention, we simplify the notation of SpSG operations into the form $\{U || \{R | \mathbf{t}\}\}$, where $U$ represents the proper or improper rotation $\theta R_S$ indicated in equation (\ref{e-SpSGtransf-spin}), and therefore, if $U$ is an improper rotation, the whole operation includes time-reversal, and this inclusion not only applies to the spins but also to the orbital degrees of freedom in equation (\ref{e-SpSGtransf-orbital}) and any other time-related variables in the system, like momenta. Hence, while the symbol $\{U || \{R | \mathbf{t}\}\}$ denotes the space operation part as $\{R | \mathbf{t}\}$, it is important to take into account that this space operation may include time-reversal, depending on the value of the determinant of $U$, although it is not explicitly indicated.
	
	In the case of an experimental magnetic structure, equation (\ref{e-SpSGtransf-orbital}) about the orbital magnetic moments is difficult to assess, as orbital and spin contributions generally remain unresolved. Given the expected smallness or null value of the orbital contribution, equation (\ref{e-SpSGtransf-spin}) is usually assumed to be applicable to the determined atomic magnetic moments \cite{Chen2024, Jiang2024,Xiao2024}. However, it should be noted that this assumption may fail, and equations (\ref{e-SpSGtransf-orbital}) and (\ref{e-SpSGtransf-spin}) imply that under the constraints of an SpSG (and therefore assuming negligible SOC), orbital atomic magnetic moments and spin moments may be forced to have different directions. This can only happen in the case of non-coplanar magnetic structures because, as explained below, the SpSGs of collinear and coplanar structures forbid, through equation (\ref{e-SpSGtransf-orbital}), any magnetic ordering of orbital type \cite{Watanabe2024}. 	
	\subsection{Subgroups of SpSGs.The Nontrivial SpSG and the Spin-Only Subgroup}
	\label{s-SpSG-subgroups}
	Several important subgroups can be distinguished in an SpSG. The \emph{spin-only} subgroup is formed by the operations of type $\{ U || \{1|0\} \}$ , i.e., operations that do not involve any space operation, except the identity, or time reversal in the case that $\det(U) = -1$. Following the notation of \citet{Chen2024}, if we call \Gso~the spin-only subgroup, the full SpSG, say \Gss, can be described as the direct product of a so-called non-trivial SpSG, \Gnt, and the \emph{spin-only} subgroup \Gso~\cite{Litvin1974}:
	\begin{equation}
		\label{e-Gss-descom}
		\textrm{G}_{\textrm{\footnotesize{SS}}} = \textrm{G}_{\textrm{\footnotesize{NT}}} \times \textrm{G}_{\textrm{\footnotesize{SO}}}.
	\end{equation}
	
	Note that, by definition, each space operation $\{ R|{\bf t}\} $ in \Gnt~is paired with one, and only one, spin operation $ U $.
	
	Only the SpSGs of collinear and planar structures have \emph{spin-only} subgroups \Gso~different from the trivial identity. Collinear structures have all the same \Gso, formed by the continuous point group of all rotations around the direction of the spins and all mirror planes containing this direction. In a similar form as in \citet{Chen2024}, we will design this spin-only subgroup, common to all collinear structures as \ssgcol. Although formally in an SpSG the collinearity direction is arbitrary with respect to the crystal lattice, for reasons explained below, we also indicate explicitly a specific orientation with respect to the lattice of the operations by means of a subscript $ {\bf n} $.
	
	The spin-only subgroup \Gso~of all coplanar structures is formed by the identity and a mirror plane with the orientation of the spin planes, i.e., $ \{ m_{\bf n} || \{1|0\} \} $, with $ {\bf n} $ indicating the perpendicular direction to the spin planes. In an analogous manner to the collinear \Gso, we denote the group as \ssgcop, where ${\bf n}$ indicates a specific direction with respect to the lattice. Equation (\ref{e-Gss-descom}) implies that collinear and coplanar structures have very specific SpSGs, distinguishable by their spin-only subgroup, either \ssgcol~or \ssgcop. We shall call them \emph{collinear} and \emph{coplanar} SpSGs, respectively. It is important to stress that this formally implies that collinearity and coplanarity are always \emph{symmetry-protected} in a SOC-free structure. We shall call all other SpSGs, which have as \Gso~only the identity, \emph{non-coplanar} SpSGs, since they can only be associated with magnetic structures that are neither collinear nor coplanar.
	
	In collinear and coplanar SpSGs, their nontrivial subgroup \Gnt~defined by equation (\ref{e-Gss-descom}) is not unique. Keeping the group structure, the $ U $ of some of the operations $ \{ U \| \{R|{\bf t}\} \} $ of \Gnt~can be substituted by its product with some spin operation of the corresponding spin-only subgroup \ssgcol~or \ssgcop. In the case of collinear structures, the nontrivial \Gnt~is usually chosen such that the $ U $ operations are either the identity or the inversion. This can always be done because all possible $ U $ operations compatible with collinearity (i.e., arbitrary proper or improper rotations about the spin direction $ {\bf n} $, 2-fold axes perpendicular to $ {\bf n} $, or planes containing $ {\bf n} $) can be written as product of a $ U $-operation of \Gso~and the identity or the inversion. Thus, the \Gnt~of a collinear SpSG is assimilable to a Shubnikov-like group, where each space operation is completed with a spin operation $ +1 $ or $ -1 $, in a way similar to what is done with ordinary MSGs. However, since in the SpSG the space operations do not act on the spins, this Shubnikov-like nontrivial SpSG is generally different from the MSG of the structure. While the nontrivial \Gnt~of a collinear SpSG, defined by a Shubnikov-like group, is independent of the spin-lattice orientation, the MSG, which is also a subgroup of the SpSG and is also described by a Shubnikov group, generally depends on the direction of the spins with respect to the crystal structure. Several examples of this situation will be discussed below.
	
	In the case of coplanar SpSGs, by convention the nontrivial groups \Gnt~are chosen such that their spin operations $ U $ are all proper rotations in 3D. This choice can always be made \cite{Litvin1974}, since any improper $ U $-operation can be automatically transformed into a proper one by multiplying it by a mirror operation of \Gso. In the case of non-coplanar SpSGs, the spin-only group is trivial, and the full SpSG coincides with the nontrivial subgroup.
	
	 Another important subgroup of an SpSG is formed by all operations of type $ \{ 1 \| \{ R, {\bf t} \} \} $, i.e., space operations that are not accompanied by any spin rotation, nor by time reversal. By definition, this is a subgroup of the nontrivial subgroup of the SpSG. The set of space operations $ \{ R|{\bf t} \} $ of this subgroup is an ordinary space group, say L$_0 $. If we call G$_0 $ the ordinary space group formed by all space operations $ \{ R|{\bf t} \} $ present in \Gnt, this space group  G$_0$  can then be decomposed in cosets with respect to L$_0 $:
	\begin{displaymath}
		\textrm{G}_0 = \textrm{L}_0 + g_2 \textrm{L}_0 + \dots + g_n \textrm{L}_0.
	\end{displaymath}
	As L$_0 $ is a normal subgroup of G$_0 $ \cite{Litvin1974}, the cosets in the above equation form a factor group G$_0/$L$_0$ with coset representatives $ \{ g_i \} $. All space operations in a coset $ g_i $L$_0 $ have associated the same spin point-group operation, say $ U_i $. Hence, the point-group formed by all spin operations $ \{ U_i \} $ present in \Gnt~are isomorphic to the factor group G$_0/$L$_0$. This is a property that has been systematically applied for the enumeration of non-trivial SpSGs \cite{Chen2024,Jiang2024}.
	
	The mentioned recent works that classify and enumerate SpSGs use different alternative notations, and the establishment of a unified nomenclature will require still time and effort. We will therefore not enter into notation details in this work, and when describing a specific SpSG, we will indicate its symbol in the notation proposed by \citet{Chen2024}, complemented with a full description of a set of generators of the group, if necessary. These authors also use a four-index notation, $ N_1.N_2.i_k.n_1 $, for the nontrivial part of the SpSGs, where $ N_1 $ and $ N_2 $ are the numerical indices in the International Tables for Crystallography \cite{Aroyo2016} for the space groups L$_0 $ and G$_0 $, respectively, associated with the SpSG. The number $ i_k $ is the \emph{klassengleich} index of L$_0 $ with respect to G$_0 $, and $ n_1 $ is just an ordering index. The klassengleich index $ i_k $ indicates the multiplication factor of a primitive unit cell describing the lattice of L$_0 $ with respect to that of G$_0 $. Therefore, if $ i_k > 1 $, the nontrivial subgroup \Gnt~of the SpSG necessarily includes some operations of type $ \{ U \| \{1 | {\bf t}\} \} $, which are very important when considering the corresponding SpPG.
	
	In an SpSG, by definition, the spin operations $ U $ are independent of the space operations. This has led to the convention of using an orthonormal reference system for the description of these operations, fully independent of the crystallographic axes, with its orientation only partially fixed in collinear and coplanar structures to the spin directions or the spin planes, respectively, and with an arbitrary orientation with respect to the lattice. However, we have here a situation similar to that of ordinary space groups, where the arbitrariness of the origin in space is not an obstacle to fix this origin at convenience. In the same way, in the SpSG formalism, the arbitrariness of the global orientation of the spin system with respect to the lattice should not be an obstacle to choose and fix a convenient reference frame for the spin system with respect to the lattice. In our view, in most cases it is convenient to choose this frame equal to that for the space operations. In this work, we will then express the operations $ U $ and $ R $ of any operation $ \{ U \| \{ R | {\bf t} \} \} $ in a common reference system defined by the conventional unit cell and the crystallographic axes which are normally used for the description of the space operations. This does not imply any loss of generality, as an arbitrary global orientation of the $ U $ operations with respect to the crystallographic axes can always be introduced if desired, when describing these operations in the chosen reference frame.
	
	In addition, in most cases magnetic anisotropy cannot be fully ignored and spins have a very specific relative orientation with respect to the crystallographic axes. Even with a hypothetical null SOC and the energy being independent of the relative orientation of spin and space operations, magnetic crystal tensor properties are measured and quantified in a reference frame locked to the crystal structure, and therefore their symmetry-adapted form in this reference frame depends in general on the relative orientation of the spin and space operations.  Therefore, for practical reasons, when dealing with the SpSG of a specific structure, the spin operations in the SpSG will be described (locked) under the specific spin-lattice orientation observed in the structure. As shown below, this allows a consistent comparison of the SpSG and MSG symmetries that can be assigned to the structure, and their corresponding constraints.
	
 \subsection{Spin Point Groups}
\label{s-SpPGs}	
	For the symmetry properties of crystal tensors, only the SpPG is relevant. This is formed by the pairs of point-group operations $\{U || R\}$ present in the SpSG operations. The subgroup of operations $\{U || 1\}$ form the \emph{spin-only} point group \Pso. Similarly to equation (\ref{e-Gss-descom}), the full SpPG can be decomposed in a direct product of a so-called “nontrivial” SpPG, \Pnt, and the spin-only point group \Pso:
	\begin{equation}
	\label{e-Pss-descom}
	\textrm{P}_{\textrm{\footnotesize{S}}} = \textrm{P}_{\textrm{\footnotesize{NT}}} \times \textrm{P}_{\textrm{\footnotesize{SO}}}
	\end{equation}
	
	However, the similarity with equation (\ref{e-Gss-descom}) may be misleading because, as discussed above, \Gnt~may have operations of type $\{U \| \{1 \, | \, {\bf t}\}\}$, with ${\bf t}$ being not a lattice translation. These operations form the so-called spin-translation group \Gst, and their point group operations $\{U \| 1\}$ will belong to \Pso. Hence, the SpPGs \Pnt~and \Pso~do not necessarily coincide with the point groups separately associated with \Gnt~and \Gso, \Pso~being in general a supergroup of the point group associated with \Gso. This means that while there are only two possible \emph{spin-only} space groups \Gso, associated with collinear and coplanar structures, the number of possible \emph{spin-only} point groups \Pso~does not have this restriction and may also be relevant for non-coplanar SpSGs.
	
	The spin-only subgroup \Pso~in equation (\ref{e-Pss-descom}) can then generally be decomposed in the direct product of two subgroups:
	\begin{equation}
	\label{e-Pso-descom}
	\textrm{P}_{\textrm{\footnotesize{SO}}} = \textrm{P}_{\textrm{\footnotesize{SOG}}} \times \textrm{P}_{\textrm{\footnotesize{SOintr}}},
	\end{equation}
	where \Psoin~is the intrinsic (or trivial) point group \ssgcol~or \ssgcop~present in collinear and coplanar SpSGs, and \Psog~is the spin-only point group that may be present in the nontrivial \Gnt. The additional \Psog~must be considered only in the case that the \textit{klassengleich} index $i_k$ of the subgroup L$_0$ with respect to G$_0$, mentioned above, is larger than one, such that the translation lattice of L$_0$ is a sublattice of the lattice in G$_0$. For $i_k = 1$ and a non-coplanar SpSG, the SpPG is directly a nontrivial SpPG, and no spin-only subgroup must be considered. In \citet{Liu2022} an enumeration of collinear and coplanar SpPGs was carried out by restricting \Psog\, to be the identity. These groups would be valid if $i_k=1$ and could be useful for local symmetry studies even if $i_k>1$. In our study of macroscopic properties, however, \Psog\, must necessarily be included.
	
	The fact that, in contrast with SpSGs, the spin-only point subgroups \Pso~are not limited to two, and are not generally trivial, means that the term “nontrivial” assigned to the point group \Pnt~in equation (\ref{e-Pss-descom}) is somehow ill-founded. We however stick to this terminology. A derivation of the possible non-equivalent nontrivial SpPGs \Pnt~in equation (\ref{e-Pss-descom}) was done by \citet{Litvin1977}, and a total of 598 were enumerated. This derivation was done taking into account that the point-group operations $U$ can only be crystallographic.
	
	The structure of the SpPG described in equation (\ref{e-Pss-descom}) allows the derivation of the symmetry-adapted form of any tensor in a stepwise form, considering first the constraints caused by the nontrivial group \Pnt~and then adding those coming from \Pso. In many cases, \Pnt~can be chosen to coincide with the actual MPG of the structure, and \Pso~is only the intrinsic spin-only subgroup, associated with the collinearity or the coplanarity of the structure (see Section \ref{s-relation-spin-magnetic}). In such cases, the SpPG form of the tensor can be then obtained by just adding the constraints due to \Psoin~to those under the MPG of the structure.

\section{Relation between spin and magnetic groups}
	\label{s-relation-spin-magnetic}
	By definition, the SpSG of a magnetic structure does not depend on the global orientation of the spin system with respect to the lattice. However, if spin and space operations are described in the same reference frame, the subgroup of operations $\{U || \{R|{\bf t}\}\}$ that fulfill $U = +R$ or $-R$ constitute according to equations (\ref{e-Mmoment}) and (\ref{e-SpSGtransf-spin}) an MSG, which proves to be the MSG of the structure if (and only if) the SpSG is being described under the specific relative spin-lattice orientation observed in the structure. Only under this condition do the SpSG and the actual MSG of the magnetic structure have a group-subgroup relation. Conversely, the same SpSG can have different MSGs as subgroups depending on the chosen orientation of the spin operations $U$ with respect to the lattice, and as a consequence, the same SpSG can be associated with magnetic structures that have very different MSGs.
	
	Therefore, the application of the SpSG symmetry on a magnetic structure and its comparison with its MSG requires to fix a specific orientation of the spin operations $U$ with respect to the lattice, which must be consistent with the spin-lattice orientation observed in the structure. In the following, if no indication on the contrary is given, the SpSG and the SpPG of a magnetic structure will be described fulfilling this condition. In this way, the stronger symmetry constraints on the tensors under its SpPG can be compared with those expected under the MSG, when SOC effects are taken into account. This is consistent with the fact that in experimental magnetic structures the spins have a specific global orientation (and domain-related ones) with respect to the lattice, as magnetic anisotropy is generally present in some form. In axial symmetric or pseudo-symmetric systems the spin orientation on the basal plane often remains undetermined, but in most cases, it is rather an experimental problem more than a physical one.
	
	We distinguish two types of experimental magnetic structures, depending on their SpSG-MSG group-subgroup relation, namely structures with minimal SpSG and structures with non-minimal SpSG.
	\begin{enumerate}
	\item Magnetic structures with minimal SpSG.
	
	In these structures both their SpSG and their MSG have the same space operations. A majority of the observed commensurate magnetic structures enter into this group. A necessary condition for this to happen is that the klassengleich index $i_k$ of the SpSG, described in Section \ref{s-SpSG-subgroups}, is either 1 or 2, as if $i_k > 2$ the SpSG must include some operations of type $\{U || \{1|{\bf t}\}\}$ with $U \neq \pm 1$, whose space operations (namely translations) cannot be present in the MSG. Thus, $i_k \leq 2$ is required to ensure that the spin-only point group of the SpSG of these structures is limited to \Psoin~plus the additional time reversal operation, $\{-1 || 1\}$, in the case of $i_k = 2$.
	
	The only difference between the SpSG and the MSG of a structure with minimal SpSG is the intrinsic spin-only subgroup in the case of collinear and coplanar structures, while in non-coplanar structures, both groups fully coincide. Therefore, for non-coplanar structures of this type, SpSG symmetry considerations do not add any additional constraint on their material tensors. However, in collinear and coplanar structures with minimal SpSG, the spin-only subgroup makes a difference. Their SpSG can be expressed as the direct product of the actual MSG of the structure with the corresponding collinear or coplanar spin-only group, and the corresponding point groups will satisfy similar relations, namely:
	\begin{eqnarray}
		\label{e-pg-desc-col}
		\textrm{P}_\textrm{\footnotesize{S}} &=& \textrm{P}_\textrm{\footnotesize{M}} \times\, \ssgcolt\\
		\label{e-pg-desc-cop}
		\textrm{P}_\textrm{\footnotesize{S}} &=& \textrm{P}_\textrm{\footnotesize{M}} \times\, \ssgcopt
	\end{eqnarray}
	
	where \Ps~and \Pm~are the SpPG and MPG of the structure and ${\bf n}$ defines the orientation of the collinear or coplanar arrangement, as discussed above. As shown below with some examples, this implies that the symmetry-adapted form of any spin-related tensor for these structures under the SpPG can be simply derived taking the tensor form under the MPG, obtained applying the usual known rules, as can be obtained for instance in MTENSOR \cite{Gallego2019}, and then introduce the additional constraints resulting from the extra symmetry represented by \Psoin.
	
	Magnetic structures with minimal SpSG can be easily identified comparing their MSG label in the OG notation \cite{Campbell2022} with the four-index label of the nontrivial subgroup of their SpSG in the notation of \citet{Chen2024}. The space group, denoted G$_0$ in Section \ref{s-SpSG-subgroups}, formed by the space operations $\{R|{\bf t}\}$ of the nontrivial SpSG, must coincide with the space group associated with the MSG, which is formed by all its operations, disregarding the inclusion or not of time reversal. This latter space group is called the family space group F of the MSG \cite{Litvin2013,Campbell2022}. Therefore, magnetic structures with minimal SpSG fulfill that $\textrm{F} = \textrm{G}_0$. The space group type of F is given by the first number of the numerical label of the MSG in the OG notation (using the space group numerical indices of the International Tables of Crystallography), while the second number in the four-index notation of \citet{Chen2024} corresponds to G$_0$. If these two numbers coincide, and $i_k \leq 2$, G$_0$ and F necessarily coincide. The two space groups are not only of the same type, but because of the restriction on the $i_k$ value, they must be the same space group, and the structure has a minimal SpSG.
	
	From the approximately 2,000 entries of commensurate magnetic structures in the MAGNDATA database \cite{Gallego2016} about 1,500 have minimal SpSGs. We can therefore infer that in approximately 75\% of the cases the differences in the material tensor forms when considering MPG or SpPG symmetries are limited to the additional constraints coming from \Psoin~in the case of collinear and coplanar structures.
	
	\item Magnetic structures with non-minimal SpSG.
	
	These are the structures where their SpSG includes space operations that are not present in their MSG. About 25\% of the commensurate structures in MAGNDATA have non-minimal SpSGs, with their G$_0$ being a strict supergroup of F: $\textrm{G}_0 > \textrm{F}$. The klassengleich index $i_k$ of the nontrivial SpSG being larger than 2 is a sufficient condition for this strict group-subgroup relation to be satisfied, but it can also happen for $i_k = 1$ or 2. In such structures, it is clear that the additional SpPG symmetry constraints cannot be reduced to those coming just from \Psoin, because the point group of the nontrivial SpSG will be a strict supergroup of the MPG. By definition, the space group operations in G$_0$ must keep the positional crystal structure invariant. Therefore, G$_0$ can only be a strict supergroup of F if the magnetic ordering is such that the space group F associated with the MSG loses some of the space group operations of the paramagnetic phase. If we call G$_p$ the space group of the paramagnetic phase, we have then that in general for this second type of commensurate magnetic structures $\textrm{G}_p \geq \textrm{G}_0 > \textrm{F}$. Whether or not $\textrm{G}_0$ is a strict subgroup of \Gp\, makes no difference when it comes to reducing crystal tensors. We will see examples of both situations later.
	
	It will be shown below in detail that there are tensors, such as those involving only space degrees of freedom, or those involving orbital degrees of freedom, where only the space parts $R$ of the operations of the SpPG are relevant for their transformation properties. The symmetry-adapted form of these tensors under an SpPG can therefore be derived considering only the space operations in the SpPG, as done in ordinary MPGs. In the case of orbital-related tensors, one has also to consider if the operation includes time reversal or not, but the specific spin operation $U$ is irrelevant. It is therefore convenient to define, for a given SpPG, an \emph{auxiliary} ordinary MPG that we denote as the \emph{effective} MPG, \mpgeff, which can be used instead of the full SpPG to derive the symmetry-adapted form of these non-magnetic tensors or orbital-related tensors. The \mpgeff~is constructed by taking the space part $R$ of each $\{U || R\}$ operation of the SpPG, without time reversal ($R$) or with time reversal ($R'$), depending on $\det(U)$ being $+1$ or $-1$, respectively. The \mpgeff~is, in general, a supergroup of the actual MPG of the structure. The symmetry constraints under the SpPG on the mentioned type of tensors can then be obtained by considering this \mpgeff~instead of the real MPG, when applying the well-known rules for MPGs \cite{Gallego2019}. The \mpgeff~of collinear and coplanar structures is just the gray point group resulting from adding the time reversal operation to the point group of the space group G$_0$ associated with the SpSG. This is because in both collinear and coplanar structures their spin-only group, \Psoin, includes at least an operation $\{U || 1\}$ with $\det(U) = -1$, and therefore the corresponding \mpgeff~contains the time reversal operation. Thus, if P$_0$ is the point group of G$_0$, the corresponding $\textrm{MPG}_\textrm{eff}$ can be expressed as P$_0.1'$. Only if the structure has a non-minimal SpSG, this gray point group $\textrm{MPG}_\textrm{eff}$ will include point-group operations $R$ that are not present in its actual MPG.
\end{enumerate} \section{Tensor transformations under spin point group operations}
\label{s-tensor-transformations}
Given a physical property represented by a tensor $A$, the symmetry restrictions that a SpSG forces on $A$ can be found by knowing the way in which the operations $\{U||R\}$ of the SpPG transform that tensor. According to the Neumann Principle generalized to SpSGs, the operations of the SpPG on the tensor must leave it invariant, i.e., we can symbolically write $\{U || R\} A =A$.

The specific action of an operation $\{U || R\}$ depends greatly on the nature of the tensor considered. This complexity, which is already found when trying to reduce tensors according to the MPGs, \cite{Birss1963,Grimmer1993,Grimmer1994,Kleiner1966,Cracknell1973,Shtrikman1965,Kopsky2015} is higher when dealing with the SpPGs. We will begin our discussion by considering the action of $\{U || R\}$ on various tensors of rank 1, starting with examples where such action is simple and direct. These cases are those in which only the $R$-part or only the $U$-part is involved in the transformation.
 \subsection{Pure-lattice and pure-spin vectors}
	\label{s-pure-lattice-spin}
	Pure-lattice and pure-spin vectors are tensors of rank 1 whose transformations only involve either the space part $R_{ij}$ or the spin part $U_{ij}$ ($i,j=1,2,3$) of the SpPG transformation. An example of a pure-lattice vector is the electric polarization $P_i$, and an example of a pure-spin vector is the spin component of the magnetization $M_i$. The transformations in these cases have the familiar forms
	\begin{eqnarray}
		\label{e-polarization-transf}
		P_i'&=&R_{ij}P_j\\
		\label{e-Mmoment-transf}
		M_i'&=&U_{ij}M_j
	\end{eqnarray}
	
	Note that given the definition of $U$ explained in Section \ref{s-spin-groups}, it is not necessary in equation (\ref{e-Mmoment-transf}) to multiply the right-hand side by the determinant of $U$ even though ${\bf M}$ is an axial vector. 
	
	The possible orbital contribution to the magnetization is not included in equation (\ref{e-Mmoment-transf}) and will be ignored for the moment. This will be incorporated later in our treatment.
	
	The two quantities ${\bf P}$ and ${\bf M}$ are prototypes of two of the four basic ferroic effects. These four effects are rank 1 tensors which differ from each other by their specific transformations under the space inversion $\overline{1}=\{1||\overline{1}\}$ and time reversal $1'=\{-1||1\}$. They are key to analyze the action of $\{U || R\}$ on the various tensor quantities, and we will assign them different labels (V, $e$V, M, T), which specify the four different behaviors shown in Table \ref{t-ferroictensors}.

	\begin{table}[ht]
	\centering
	\caption{\label{t-ferroictensors} Transformation of the four basic ferroic effects under the space inversion and time reversal operations. The four effects are denoted by the symbols V, $e$V, M, and T. Effects V and T are odd for the space inversion, while $e$V and M are even. For the time reversal V and $e$V are even, while M and T are odd.}
	\begin{tabular}{|c|c|c|c|c|}
		\hline
		& V & $e$V & M & T \\
		\hline
		$\overline{1}=\{1\parallel\overline{1}\}$ & -1 & 1 & 1 & -1 \\
		\hline
		$1'=\{-1\parallel1\}$ & 1 & 1 & -1 & -1 \\
		\hline
	\end{tabular}
\end{table}
 			
	Thus, with reference to this table we say that ${\bf P}$ is a tensor of type V (polar Vector), and ${\bf M}$ is a tensor of type M (axial Magnetic vector). The prototypes of the other two basic effects are the moment of the polarization ${\bf A}={\bf r} \times {\bf P}$ ($e$V, axial pure-lattice vector) and the moment of the magnetization or Toroidic moment ${\bf T}={\bf r} \times {\bf M}$ (T, polar mixed vector).
	
	The transformation of a vector of type $e$V under $\{U || R\}$ is also simple,
	\begin{equation}
		\label{e-transf-eV}
		A_i'=\det(R)R_{ij}A_j
	\end{equation}
	
	The simplicity of equation (\ref{e-transf-eV}) comes from the fact that both ${\bf r}$ and ${\bf P}$ are pure-lattice vectors, and in their transformation only the space part $R$ of the operation intervenes. This is, however, not the case for a tensor of type T, which involves both space and spin operations $R$ and $U$, and whose analysis will be postponed after the discussion of the transformations of the magnetoelectric tensor.

 \subsection{The magnetoelectric tensor}
\label{s-magnetoelectric}
The magnetoelectric effect is described by a tensor of rank 2 that describes either the magnetization induced by an applied electric field $\mathbf{E}$ (inverse effect, $M_i=\alpha^{inv}_{ij}E_j$) or the polarization induced by an applied magnetic field $\mathbf{H}$ (direct effect, $P_i=\alpha^{dir}_{ij}H_j$). 

Since $\mathbf{E}$ is a pure-lattice vector and $\alpha^{inv}_{ij}$ must transform as the product $M_i E_j$, we easily obtain the transformation law of the inverse effect,
\begin{equation}
	\label{e-magnetoelectric-transf-inv}
	\alpha^{inv}_{ij}\,' = U_{ik} R_{j\ell} \alpha^{inv}_{k\ell} ,
\end{equation}
i.e., tensor $\alpha^{inv}_{ij}$ is a tensor of rank 2, whose transformation mode involves $R$ and $U$. We say that $\alpha^{inv}$ is a tensor of type MV.

The direct effect can be analyzed similarly. Thermodynamic arguments indicate \cite{Nye1985} that the tensor of the direct effect is equal to the transpose of the tensor of the inverse effect $\left[\alpha^{inv}=(\alpha^{dir})^T\right]$, so taking the transpose of equation (\ref{e-magnetoelectric-transf-inv}) we have
\begin{equation}
	\label{e-magnetoelectric-transf}
	\alpha^{dir}_{ij}\,' = R_{ik} U_{j\ell} \alpha^{dir}_{k\ell}.
\end{equation}
In the following we will use the symbols $\alpha=\alpha^{dir}$ and $\alpha^T=\alpha^{inv}$ for the direct and inverse effects, respectively. \subsection{The toroidic moment}
\label{s-toroidic}
The transformation law for the toroidic moment $\mathbf{T}$ can be deduced by noting that this quantity transforms just like the antisymmetric part of the magnetoelectric tensor (direct or inverse effect indistinctly) \cite{Spaldin2008}. This can be deduced by noticing that the quantities $\alpha^T_{ij}$ (we take the inverse effect as an example) transform as the product $M_i E_j$, so that $(\alpha^T_{ij}-\alpha^T_{ji})$ will transform as $M_i E_j - M_j E_i$. Since the electric field transforms as the position vector $\mathbf{r}$, then $M_i E_j - M_j E_i$ will transform as the $k$-component of the vector $\mathbf{M} \times \mathbf{r}$, that corresponds to the association $(i=1, j=2) \rightarrow k=3$ and circular permutations.

The components $T_i$ can therefore be assimilated to the quantities $\frac{1}{2}\varepsilon_{ijk} \alpha^T_{jk}$ from the point of view of their transformation laws, where $\varepsilon_{ijk}$ is the Levi-Civita symbol. We can say that $\mathbf{T}$ is a quantity of type \{MV\} (or \{VM\}), denoting the curly brackets the antisymmetric part. If we define a tensor $\oal_{ij} = 2M_ix_j$ (where $x_j$ are the components of $\mathbf{r}$), then we directly have $T_i=\frac{1}{2}\varepsilon_{ijk}\oal_{jk}$. Writing this tensor as a sum of a symmetric part $\oal^s$ and an antisymmetric part $\oal^a$, i.e., $\oal=\oal^s+\oal^a$, with $\oal^s=\frac{1}{2}(\oal_{ij}+\oal_{ji})=M_ix_j+M_jx_i$, and $\oal^a=\frac{1}{2}(\oal_{ij}-\oal_{ji})=M_ix_j-M_jx_i$ we will have from equation (\ref{e-magnetoelectric-transf-inv}),
\begin{eqnarray}
	\label{e-oal-symmetric}
	\oal^a_{ij}\,'&=&\frac{1}{2}(U_{im}R_{j\ell}-U_{jm}R_{i\ell})(\oal^a_{m\ell}+\oal^s_{m\ell})\\
	\label{e-oal-antisymmetric}
	\oal^s_{ij}\,'&=&\frac{1}{2}(U_{im}R_{j\ell}+U_{jm}R_{i\ell})(\oal^a_{m\ell}+\oal^s_{m\ell})
\end{eqnarray}

from which we can deduce the transformation law for $\mathbf{T}$. It is interesting to note that equations (\ref{e-oal-symmetric}) and (\ref{e-oal-antisymmetric}) indicate that the transformations for $\oal^a$ and $\oal^s$ are, in general, coupled. In other words, these transformations cannot be written in the usual form $\oal^a_{ij}\,'=X_{imj\ell}\oal^a_{m\ell}$ or $\oal^s_{ij}\,'=Y_{imj\ell}\oal^s_{m\ell}$ (with $X$, $Y$ suitable transformation matrices), because in the right-hand sides of equations (\ref{e-oal-symmetric}) and (\ref{e-oal-antisymmetric}) there are also contributions dependent on $\oal^s$ and $\oal^a$, respectively. This means that neither $\oal^s$ nor $\oal^a$ (and thus the toroidic moment) are true tensors for SpPG transformations. From equations (\ref{e-oal-symmetric}) or (\ref{e-oal-antisymmetric}) it can be deduced that $\oal^s$ and $\oal^a$ become uncoupled if 
\begin{displaymath}
	U_{im}R_{j\ell}-U_{jm}R_{i\ell}+U_{i\ell}R_{jm}-U_{j\ell}R_{im}=0
\end{displaymath}
and
\begin{displaymath}
	U_{im}R_{j\ell}-U_{i\ell}R_{jm}+U_{jm}R_{i\ell}-U_{j\ell}R_{im}=0
\end{displaymath}
i.e.,
\begin{equation}
	\label{e-oal-restriction}
	U_{im}R_{j\ell}-U_{j\ell}R_{im}\pm(-U_{jm}R_{i\ell}+U_{i\ell}R_{jm})=0
\end{equation}
which, in general, is not satisfied. A special case occurs for the MPG operations, in which $U_{\ell n} = \pm R_{\ell n}$. When this condition is met, it can be easily seen that equation (\ref{e-oal-restriction}) does certainly hold. 

Consequently, to obtain the symmetry-adapted form of $\mathbf{T}$, we must first consider the symmetry invariance of a tensor $\oal$ which transforms similarly to the magnetoelectric tensor, using equation (\ref{e-magnetoelectric-transf-inv}) or (\ref{e-magnetoelectric-transf}), and then take its antisymmetric part by means of the product of this tensor with the Levi-Civita tensor. Note that this procedure does not require the use of equations (\ref{e-oal-symmetric}) and (\ref{e-oal-antisymmetric}). Thus, the component $T_p'$ of the toroidic moment transformed by the operation $\{U || R\}$ will be given by 
\begin{equation}
	\label{e-toroidic-moment}
	T_p' =\frac{1}{2}\varepsilon_{pij} U_{ik}R_{j\ell} \oal_{k\ell}.
\end{equation}

The complexity of this transformation law is a characteristic of SpPGs and leads to more laborious tensor symmetry-reductions than those for MPGs.
 \subsection{Equilibrium properties}
\label{s-equilibrium-properties}
Once the transformation properties of the four basic ferroic effects have been deduced, we can obtain the corresponding transformations for the different equilibrium properties through their constitutive equations. They can be described in each case by an appropriate combination of the labels V and M, which accounts for the intrinsic symmetry of the tensor. These combinations constitute symbols that generalize the so-called Jahn symbols \cite{Gallego2019,Jahn1949} used with the MPGs.

Table \ref{t-equilibrium-properties} lists a selection of equilibrium properties, the constitutive equation, the Jahn symbols for the MPGs and SpPGs, and an outline of the transformation law in the case of the SpPGs. The table only lists tensors of properties where spin magnetism is involved, and therefore their transformation rules have to be modified when considering SpPG symmetry. In the case of pure-lattice tensors, the known transformation rules for ordinary space group operations are still in place, as they only involve space operations $R$. Hence, in tensors such as electric polarization, dielectric susceptibility, or piezoelectric tensor, a difference between the constraints when considering MPG and SpPG symmetry can only occur in structures with a non-minimal SpSG, where the space group G$_0$ associated with the SpSG is a supergroup of the family group F of its MSG (see Section \ref{s-relation-spin-magnetic}). The calculation of the symmetry-adapted form of these tensors under the SpPG can be obtained by applying the well-known transformation rules for MPGs under the symmetry given by \mpgeff, which was defined in Section \ref{s-relation-spin-magnetic}.

\begin{table}[ht]
	\centering
	\caption{\label{t-equilibrium-properties}Selection of some equilibrium properties with their Jahn symbols for the MPGs and SpPGs and their transformation laws under the SpPG. Only tensors related with spin magnetism are listed (see text). $\varepsilon_{ijk}$ is the Levi-Civita symbol, and $\varepsilon_{jk}$ and  $\sigma_{jk}$ stand for the strain and stress tensors respectively. In the case of MPGs the label $e$ in the Jahn symbol indicates an axial tensor and the label $a$ a magnetic tensor, i.e., odd for time reversal. This means that the law of tensor transformation includes a change of sign for improper operations ($e$) or for operations that include time reversal ($a$). The square brackets and curly brackets indicate symmetry and antisymmetry of pairs of indices respectively. The symmetric or antisymmetric character is not explicit in the outline of the transformation law indicated in last column.}
	\begin{tabular}{|l|l|l|l|}
		\hline
		\textbf{Tensor Description} & \textbf{Defining } & \textbf{Jahn Symbol} & \textbf{Transformation} \\
		 & \textbf{Equation} & \textbf{(MPG/SpPG)} & \textbf{laws (SpPG)} \\
		\hline
		Magnetization & $M_i$ & $ae$V/M & $UM$ \\
		\hline
		Polar Toroidic moment & $T_i$ & $a$V/\{MV\} & $UR\oal$; $T_i=\frac{1}{2}\varepsilon_{ijk}\oal_{jk}$ \\
		\hline
		Magnetic susceptibility & $M_i = \chi^m_{ij} H_j$ & \vsim/\msim & $UU\chi^m$ \\
		tensor $\chi^m_{ij}$ &  &  &  \\
		\hline
		Magnetoelectric tensor & $M_i = \alpha^T_{ij} E_j$ & $ae$V$^2$/MV & $UR \alpha^T$ \\
		$\alpha^T_{ij}$ (inverse effect) &  &  \\
		\hline
		Electrotoroidic tensor & $t_i = \theta_{ij} E_j$ & $a$V$^2$/\{MV\}V & $URR b$; $\theta_{ij}=\frac{1}{2}\varepsilon_{ik\ell}b_{k\ell j}$ \\
		$\theta_{ij}$ (inverse effect) &  &  & \\
		\hline
		Piezotoroidic tensor & $t_i = \gamma_{ijk} \sigma_{jk}$ & $a$V\vsim/\{MV\}\vsim & $URRR b$; $\gamma_{ijk}=\frac{1}{2}\varepsilon_{i\ell p}b_{\ell pjk}$ \\
		$\gamma_{ijk}$ (direct effect) &  &  &  \\
		\hline
		Second order magnetoelectric & $P_i = \alpha_{ijk} H_j H_k$ & V\vsim/V\msim & $RUU \alpha$ \\
		tensor $\alpha_{ijk}$ (direct effect) &  &  &  \\
		\hline
		Piezomagnetic tensor & $M_i = \Lambda_{ijk} \sigma_{jk}$ & $ae$V\vsim/ M\vsim & $URR \Lambda$ \\
		$\Lambda_{ijk}$ (direct effect) &  &  &  \\
		\hline
		Magnetostriction tensor $N_{ijk\ell}$ & $\varepsilon_{ij} = N_{ijk\ell} H_k H_\ell$ & \vsim\vsim/\vsim\msim & $RRUU N$ \\
		\hline
	\end{tabular}
\end{table}
 
Apart from the magnetization, there is in Table \ref{t-equilibrium-properties} one case (magnetic susceptibility) where the transformation includes only the $U$-part, in which the invariance against $\{U || R\}$ is written in the simple form
\begin{equation}
	\label{e-magnetic-susceptibility}
	\chi^m_{ij} = U_{ik}U_{j\ell}\chi^m_{k\ell}.
\end{equation}

In all other examples, both $R$s and $U$s are involved in various combinations. Particularly complicated are the tensor transformations whose symbol includes \{MV\}.

An extension of Table \ref{t-equilibrium-properties}, with a more comprehensive list of material properties, is given in the Supporting Information (Table S1).

As an example of how to find the symmetry-adapted shape of a given tensor, we choose the electrotoroidic effect $\theta_{ij}$ (type \{MV\}V). $\theta_{ij}$ is reduced in a two-step process. First we take a type MVV rank-3 tensor, $b_{ijk}$, and reduce it. Then, we contract the first two indices by means of the product with $\varepsilon_{ijk}$. More explicitly, first we will find the $b_{ijk}$ tensor invariant under all operations $\{U || R\}$ by requiring
\begin{equation}
	\label{e-electrotoroidic-pre}
	b_{ijk} = U_{i\ell} R_{jm} R_{kn} b_{\ell mn},
\end{equation}
and, afterwards, we will take the antisymmetric part of $b_{ijk}$ with respect to the first two indices in the form
\begin{equation}
	\label{e-electrotoroidic}
	\theta_{pk} = \frac{1}{2}\varepsilon_{pij} b_{ijk}.
\end{equation}

The additional symmetries indicated in Table \ref{t-equilibrium-properties} by the square brackets are easy to handle. For example, to reduce the piezotoroidic tensor $\gamma_{ijk}$ (direct effect) that transforms according to \{MV\}\vsim, we will first take a type MVVV auxiliary tensor of rank 4 and require its invariance under the SpPG, i.e.,
\begin{equation}
	\label{e-piezotoroidic-pre}
	b_{ijk\ell} = U_{im} R_{jn} R_{kp} R_{\ell q}b_{mnpq}.
\end{equation}
Now, once equation (\ref{e-piezotoroidic-pre}) is solved, tensor $\gamma_{ijk}$ is obtained by means of the expression
\begin{equation}
	\label{e-piezotoroidic}
	\gamma_{pk\ell} =\frac{1}{4} \varepsilon_{pij} (b_{ijk\ell} +  b_{ij\ell k}),
\end{equation}
which takes out the antisymmetric part of $b_{ijk\ell}$ in the first two indices and symmetrizes the final tensor in the $k$ and $\ell$ indices.

We end this section by noting that the Jahn symbols in Table \ref{t-equilibrium-properties} can be used not only to derive the symmetry restrictions of the tensors under a given SpPG, but they also permit to obtain the relation between tensors corresponding to two structures with the same SpPG, differing only in a global spin rotation. Thus, if this rotation (proper or improper) is described by a matrix $P$, the new tensor is obtained from the old one after substituting $U$ by $P$ and taking a rotation $R$ equal to the identity, $R_{ij} = \delta_{ij}$, in the last column of Table \ref{t-equilibrium-properties}. For example, in the case of the magnetization
\begin{equation}
	\label{e-magnetization-rotated}
	M_i' = P_{ij} M_j,
\end{equation}
where $\mathbf{M}'$ is the magnetization of the structure with the spins rotated. Similarly, the magnetic susceptibility of the rotated spin structure $\chi^m\,'$ will be
\begin{equation}
	\label{e-susceptibility-rotated}
	\chi^m_{ij}\,' = P_{ik}P_{j\ell} \chi^m_{j\ell},
\end{equation}
and in the case of the inverse magnetoelectric tensor we will have
\begin{equation}
	\label{e-magnetoelectric-rotated}
	\alpha^T_{ij}\,' = P_{ik} \alpha^T_{kj}.
\end{equation}
This is of interest, for example, for relating the tensors of two collinear (or coplanar) structures with different orientations of the spin direction (or of the spin plane) with respect to the lattice. Note, however, that their scope is wider and can be used more generally, even with non-coplanar structures.

In this respect it is interesting to point out that one could alternatively define the symmetry-adapted form of the tensors under a SpPG using two different reference frames for spin and lattice variables, so that the spin-related indices of the tensor refer to a spin reference system independent of the one used for the lattice. This approach permits to obtain a general description for the tensors under the SpPG symmetry. For example, the magnetoelectric tensor can be defined as a tensor $\alpha^T_{i'j}$ with unprimed lattice indices and primed indices referring to the spin space. Thus, the physical meaning of this coefficient is that an electric field along $j$ in the lattice induces a magnetization along $i'$, the direction $i'$ being defined with respect to the reference frame of the spins, which can be chosen totally independent of the lattice. We will return to this point later, when we analyze some examples.
 \subsection{Equilibrium properties related with orbital degrees of freedom}
\label{s-equilibrium-orbital}

The Jahn symbol of all tensors in Table \ref{t-equilibrium-properties} contains the letter “M”, as they correspond to magnetic tensor properties resulting from the electronic spins. But, in general, all these tensors may also have a contribution of orbital origin. As discussed in Section \ref{s-spin-space-groups}, contrary to the atomic spins, orbital magnetic moments are locked to the lattice even in SOC-free systems, and therefore the orbital part of these tensors transforms according to the usual Jahn symbol for MPGs. For instance, upon an operation $\{U || R\}$ of the SpPG, the magnetization $\mathbf{M}_{orb}$ of orbital origin is transformed as a magnetic axial vector according to the space operation R, incorporating the possible time reversal if $\det(U) = -1$ \cite{Watanabe2024}. Thus, we will have the counterpart of equation (\ref{e-Mmoment-transf}),
\begin{equation}
	\label{e-magnetization-orbital}
	M_{orb,i}\,' = \det(U)\det(R)R_{ij} M_{orb,j}.
\end{equation}
The associated Jahn symbol is $ae$V, as in an ordinary MPG. Note, however, that here the MPG to be used is \mpgeff, described in Section \ref{s-relation-spin-magnetic}, whose elements in the $\{U||R\}$ notation are of the form $\{\det(U)\det(R)R||R\}$.

Orbital contributions of properties listed in Table \ref{t-equilibrium-properties} are thus transformed differently from their spin contributions. For example, the magnetoelectric tensor (inverse effect) has an orbital component $\alpha^{orb\,T}$ whose transformation law corresponds to the Jahn symbol $ae$V$^2$. This means that for an operation $\{U||R\}$ the transformation is of the form
\begin{equation}
	\label{e-magnetoelectric-orbital}
	\alpha^{orb\,T}_{ij}\,' = \det(U)\det(R)R_{ik}R_{j\ell} \alpha^{orb\,T}_{k\ell}.
\end{equation}

The toroidic moment also has an orbital component $\mathbf{T}_{orb}$. Since the operations $\{\det(U)\det(R) R || R\}$ of \mpgeff~verify equation (\ref{e-oal-restriction}), the transformation law is simpler here than in the case of the spin component. For the orbital contribution we have decoupled the transformations of the symmetric and antisymmetric parts of the magnetoelectric tensor, which gives rise to the simple result
\begin{equation}
	\label{e-toroidic-orbital}
	T_{orb,i}\,' = \det(U)R_{ij} T_{orb,j}.
\end{equation}

As a last example we take the orbital part of the magnetic susceptibility, which transforms as
\begin{equation}
	\label{e-susteptibility-orbital}
	\chi^{m,orb}_{ij}\,' = R_{ik}R_{j\ell} \chi^{m,orb}_{k\ell},
\end{equation}
i.e., in the same way as in a non-magnetic crystal.

Therefore, in general, the symmetry-adapted form of the tensors of orbital origin can be simply derived using the transformation rules for the \mpgeff. The tensors for the full properties are then the sum of the tensors for the spin and orbital contributions. This has important simple consequences because, as shown in Section \ref{s-relation-spin-magnetic}, the \mpgeff~of all collinear and coplanar structures are gray. This implies that the orbital contribution to all tensors that are odd under time reversal (i.e., “$a$” present in the Jahn symbol) is necessarily null in collinear and coplanar structures, if SpSG symmetry is valid. On the other hand, for tensors that are even under time reversal (i.e., “$a$” not present in the Jahn symbol), the collinearity or coplanarity of the structure does not introduce any specific restriction to their orbital contributions. Finally, it should also be noted that the tensors accounting for the orbital contributions under the symmetry constraints of a SpPG are independent of the global orientation of the spin arrangement. 

Table S1 in the Supporting Information also shows separately the transformation rules for the orbital and spin contributions in a selection of equilibrium properties.
 \subsection{Constraints on equilibrium tensors of collinear and coplanar magnetic structures}
\label{s-equilibrium-constraints}

As indicated by equation (\ref{e-Pss-descom}), any SpPG is the direct product of a nontrivial part and a spin-only point group. Collinear and coplanar structures are characterized by the fact that they always possess a certain minimum symmetry, \Psoin, in their spin-only point group \Pso. This symmetry alone produces certain general restrictions on some tensor properties, which can be derived separately.

In collinear materials the spin operations $U$ of \Psoin~form the continuous group $\infty m$, and in the coplanar case the group $m$. In order to derive the tensor constraints on a general basis, and following the usual convention, we take the $z$-axis parallel to the spins in the case of collinear groups, and in the case of coplanar groups the plane of symmetry is taken perpendicular to $z$. Hence, the generators of the two \Pso~are $(\{\infty_z || 1\}, \{ m_x || 1 \})$ and $(\{ m_z || 1 \})$, respectively. For each specific structure, the resulting tensor constraints derived for this generic \( z \)-direction will have then to be translated to the actual collinear or coplanar orientation with respect to the lattice, which is present in the structure. These operations strongly constrain the form of some tensors as shown in Table \ref{t-equiligrium-constraints}. The table only lists tensors for spin magnetism contributions. The constraints resulting from collinearity or coplanarity in the case of tensor contributions of orbital origin were already discussed in the previous section, where they were reduced to the simple rule that tensors odd for time reversal are null, while for even ones they do not imply any specific restriction. This means that time-odd tensors in collinear and coplanar structures under SpSG symmetry can only have contributions of spin origin. For pure-lattice tensors, where only space operations are involved, obviously collinearity or coplanarity do not introduce any specific restriction, and are not included in the table either.

\begin{table}[ht]
	\centering
	\caption{\label{t-equiligrium-constraints}Constraints imposed by collinearity and coplanarity on some magnetic tensors of equilibrium properties, assuming SpPG symmetry. Only tensors related with spin magnetism are listed. The $z$ direction is taken as the spin direction in the collinear case and as the direction perpendicular to the spin planes in the coplanar case.}
	\begin{tabular}{|c|l|l|}
		\hline
		\textbf{Tensor} & \textbf{Collinear Structure} & \textbf{Coplanar Structure} \\
		\hline
		Magnetization $M_i$ & \multirow{2}{*}{$(0,0,M_3)$} & \multirow{2}{*}{$(M_1,M_2,0)$} \\
		(spin contribution) && \\
		\hline
		Toroidic moment $T_p$ & \multirow{3}{*}{$\left(\begin{array}{ccc}0&0&0\\0&0&0\\\oal_{31}&\oal_{32}&\oal_{33}\end{array}\right)$} & \multirow{3}{*}{$\left(\begin{array}{ccc}\oal_{11}&\oal_{12}&\oal_{13}\\\oal_{21}&\oal_{22}&\oal_{23}\\0&0&0\end{array}\right)$} \\
		(spin contribution) && \\
		 $T_p=\frac{1}{2}\varepsilon_{pij}\oal_{ij}$ && \\
		\hline
		Magnetic susceptibility $\chi^m_{ij}$ & \multirow{3}{*}{$\left(\begin{array}{ccc}\chi^m_{11}&0&0\\0&\chi^m_{11}&0\\0&0&\chi^m_{33}\end{array}\right)$} & \multirow{3}{*}{$\left(\begin{array}{ccc}\chi^m_{11}&\chi^m_{12}&0\\\chi^m_{12}&\chi^m_{22}&0\\0&0&\chi^m_{33}\end{array}\right)$} \\
		(spin contribution) &&\\
		 &&\\
		\hline
		Magnetoelectric tensor $\alpha^T_{ij}$ & \multirow{3}{*}{$\left(\begin{array}{ccc}0&0&0\\0&0&0\\\alpha^T_{31}&\alpha^T_{32}&\alpha^T_{33}\end{array}\right)$} & \multirow{3}{*}{$\left(\begin{array}{ccc}\alpha^T_{11}&\alpha^T_{12}&\alpha^T_{13}\\\alpha^T_{21}&\alpha^T_{22}&\alpha^T_{23}\\0&0&0\end{array}\right)$} \\
		(spin contribution) (inverse effect) &&\\
		 &&\\
		\hline
		Electrotoroidic tensor $\theta_{pk}$ & $b_{1ij}=b_{2ij}=0$, & $b_{1ij}$, $b_{2ij}$ no restriction,\\
		(spin contribution) (inverse effect) & $b_{3ij}$ no restriction& $b_{3ij}=0$ \\
		$\theta_{pk}=\frac{1}{2}\varepsilon_{pij}b_{ijk}$ &&\\
		\hline
		Piezotoroidic tensor $\gamma_{pk\ell}$ & $b_{1ijk}=b_{2ijk}=0$, & $b_{1ijk}$, $b_{2ijk}$ no restriction, \\
		(spin contribution) (direct effect) & $b_{3ijk}$ no restriction& $b_{3ijk}=0$ \\
		$\gamma_{pk\ell}=\frac{1}{2}\varepsilon_{pij}b_{ijk\ell}$ &  &  \\
		\hline
		Second order magnetoelectric & \multirow{3}{*}{$\left(\begin{array}{cccccc}\alpha_{11}&\alpha_{11}&\alpha_{13}&0&0&0\\\alpha_{21}&\alpha_{21}&\alpha_{23}&0&0&0\\\alpha_{31}&\alpha_{31}&\alpha_{33}&0&0&0\end{array}\right)$} & \multirow{3}{*}{$\left(\begin{array}{cccccc}\alpha_{11}&\alpha_{12}&\alpha_{13}&0&0&\alpha_{16}\\\alpha_{21}&\alpha_{22}&\alpha_{23}&0&0&\alpha_{26}\\\alpha_{31}&\alpha_{32}&\alpha_{33}&0&0&\alpha_{36}\end{array}\right)$} \\
		 tensor $\alpha_{ijk}$ &  &  \\
		(spin contribution) (direct effect) &  &  \\
		\hline
		Piezomagnetic tensor $\Lambda_{ijk}$ & $\Lambda_{1jk}=\Lambda_{2jk}=0$, & $\Lambda_{1jk}$, $\Lambda_{2jk}$ no restriction, \\
		(spin contribution) (direct effect) & $\Lambda_{3jk}$ no restriction & $\Lambda_{3jk}=0$ \\
		\hline
		Magnetostriction tensor $N_{ijk\ell}$ & $N_{i1} = N_{i2}$, $N_{i3}$ arbitrary,& $N_{i1}$, $N_{i2}$, $N_{i3}$, $N_{i6}$ arbitrary,\\
		($N_{ik}$ in abbreviated notation)& $N_{i4} = N_{i5} = N_{i6} = 0$; &  $N_{i4} = N_{i5} = 0$;\\
		(spin contribution) & $i=1,\ldots,6$ &  $i=1,\ldots,6$ \\
		\hline
	\end{tabular}
\end{table}
 
The constraints described in Table \ref{t-equiligrium-constraints}, resulting from the collinearity or coplanarity of the structure, i.e., from \Psoin, must be added to the symmetry-adapted form of the tensor deduced from the nontrivial subgroup of the SpPG, and the non-intrinsic spin-only group (if existing). In the case of structures with minimal SpSG (see Section \ref{s-relation-spin-magnetic}), it is sufficient to add the collinear or coplanar constraints described in the table to the symmetry-adapted form of the tensor for the actual MPG of the structure.

When dealing with properties where the spin contribution to the toroidic moment is involved, Table \ref{t-equiligrium-constraints} does not directly show the constraints due to collinearity or coplanarity. Instead, it indicates the restrictions on the tensors out of which these quantities are constructed by antisymmetrizing two of the indices. For example, for the inverse electrotoroidic effect $\theta_{ij}$ (type \{VM\}V), the form corresponding to a 3-index tensor of type VMV, $b_{ijk}$, is indicated. It is on this \emph{extended} tensor that the rest of the SpPG constraints must be applied when deducing the final form of the property in question. This issue is due to the fact that $\theta_{ij}$ is not really a genuine tensor for SpPG transformations (since $T_i$ is not, see Section \ref{s-toroidic}).

In the Supporting Information we have extended Table \ref{t-equiligrium-constraints} with more properties, and also explicitly list the restrictions on orbital contributions where applicable (Table S2).
 \subsection{Transport phenomena}
\label{s-transport}
For non-equilibrium transport properties, it is the Onsager theorem, and not the constitutive relationships, that indicates how these tensors transform under the time reversal operation \cite{Butzal1982,Eremenko1992,Grimmer1994,Shtrikman1965}. For example, it can be shown from the Onsager theorem that the electrical resistivity $\rho$, which relates electric field $\mathbf{E}$ and current density $\mathbf{J}$ ($E_i = \rho_{ij} J_j$), is transformed by time reversal in the form:

\begin{equation}
	\label{e-resistivity}
	\{-1 || 1\} \rho_{ij} = \rho_{ji}
\end{equation}

This expression allows defining a symmetric part $\rho^s$ and an antisymmetric part $\rho^a$ ($\rho = \rho^s + \rho^a$) \cite{Grimmer2017} that are even and odd for time reversal, i.e.,

\begin{equation}
	\label{e-rho-transformation}
	\{-1|| 1\} \rho^s = \rho^s, \quad \{-1 || 1\} \rho^a = -\rho^a.
\end{equation}

Therefore, since the electric field $\mathbf{E}$ is a vector of type V, we deduce that $\rho^s$ must be a tensor of type \vsim. As for $\rho^a$, it should be noted that although in principle the $U$ part of $\{U || R\}$ affects neither the electric field nor the current density, the second of equations (\ref{e-rho-transformation}) implies that there must be a sign change in the transformation if the time reversal is included in the $\{U || R\}$ operation, i.e., if $U$ is improper.  

Thus, for the symmetric part we have

\begin{equation}
	\label{e-rho-symmetric-transformation}
	\rho^s_{ij}=R_{ik}R_{j\ell}\rho^s_{k\ell}
\end{equation}

and for the antisymmetric part

\begin{equation}
	\label{e-rho-antisymmetric-transformation}
	\rho^a_{ij}=\det(U)R_{ik}R_{j\ell}\rho^a_{k\ell}
\end{equation}

In other words, $\rho^a$ is an antisymmetric magnetic tensor, whose Jahn symbol is $a$\{V$^2$\}, just as with ordinary MPGs. $\rho^s$ accounts for the ordinary electric resistivity, whereas $\rho^a$ is the responsible of the anomalous (or spontaneous) Hall effect.

Similar to the orbital components of the equilibrium properties, the transformations of $\rho^s$ and $\rho^a$ by the SpPG are formally identical to those of an MPG, and therefore, the symmetry-adapted form of the tensors can be obtained just with the methods employed for MPGs, applied to the\mpgeff~that can be associated with the SpPG. 

It is interesting to note that the restrictions imposed by the SpPGs on the magnetization and $\rho^a$ are not equivalent. This is in sharp contrast to the case of the ordinary MPGs, where it can be shown that the Jahn symbols for ${\bf M}$ and $\rho^a$ ($ae$V and $a$\{V$^2$\}, respectively) are equivalent, in such a way that the occurrence of magnetization is closely linked to the existence of the anomalous Hall effect. However, in the framework of SpPGs, the equivalence in the transformation law is between the anomalous Hall effect and just the orbital part of the magnetization. Therefore, it can be the case of having ${\bf M}_{orb}=0$ so that $\rho^a = 0$ (without SOC), and yet there is a non-zero spin component of the magnetization. Thus, there are ferromagnetic systems where the anomalous Hall effect can only be a SOC effect. Conversely, antiferromagnetic (non-coplanar) structures may exhibit an anomalous Hall effect, even with the spin macroscopic magnetization being zero \cite{Watanabe2024}. 

The application of external magnetic fields leads to the definition of new effects that are described by tensors of ranks higher than 2. For example, keeping only terms linear in ${\bf H}$,

\begin{equation}
	\label{e-rho-series}
	\rho_{ij}({\bf H})=\rho_{ij}(0)+R_{ijk}H_k+\ldots,
\end{equation}

and separating symmetric and antisymmetric parts, we have two tensors, $R^s_{ijk}=\frac{1}{2}(R_{ijk}+R_{jik})$ and $R^a_{ijk}=\frac{1}{2}(R_{ijk}-R_{jik})$, symmetric and antisymmetric in the first two indices respectively. The symmetric part of the spin component $R^s$ is of type \vsim M and accounts for the linear magneto-resistance, while the spin contribution to $R^a$ is of type $a$\{V$^2$\}M, and is the tensor describing the ordinary Hall effect \cite{Grimmer2017}. The meaning of these symbols is as follows:

\begin{equation}
	\label{e-magnetoresistance-symmetric}
	R^s_{ijk}\,'=R_{i\ell}R_{jm}U_{kn}R^s_{\ell mn}\hspace{2cm} ([\textrm{V}^2]\textrm{M})
\end{equation}

and

\begin{equation}
	\label{e-magnetoresistance-antisymmetric}
	R^a_{ijk}\,'=\det(U)R_{i\ell}R_{jm}U_{kn}R^a_{\ell mn}\hspace{2cm} (a\{\textrm{V}^2\}\textrm{M}).
\end{equation}

These transformations are also valid for the spin Hall resistivity tensor, ${\rho_{ij}}^k$, that connects the electric field with the spin current polarized in the $k$-direction ${\bf J}^k$ ($E_i={\rho_{ij}}^k{J_j}^k$) \cite{Seemann2015,Zelezny2017}. Since the relation between ${\bf J}^k$ and ${\bf J}$ is just a term that is transformed as the spin (type M), it follows that the symmetric part of ${\rho_{ij}}^k$ in $ij$ must also transform as \vsim M and the antisymmetric part as $a$\{V$^2$\}M.

Tensors $R_{ijk}$ and ${\rho_{ij}}^k$ also have orbital components (orbital Hall tensor and orbital-current Hall resistivity) \cite{Bernevig2005}. For example, in the case of the Hall effect we will have for the symmetric and antisymmetric parts the transformations

\begin{equation}
	\label{e-magnetoresistance-symmetric-orb}
	R^{orb,s}_{ijk}\,'=\det(U)\det(R)R_{i\ell}R_{jm}R_{kn}R^{orb,s}_{\ell mn}\hspace{2cm} (ae[\textrm{V}^2]\textrm{V})
\end{equation}

and

\begin{equation}
	\label{e-magnetoresistance-antisymmetric-orb}
	R^{orb,a}_{ijk}\,'=\det(R)R_{i\ell}R_{jm}R_{kn}R^{orb,a}_{\ell mn}\hspace{2cm} (e\{\textrm{V}^2\}\textrm{V})
\end{equation}

We end this section with a reference to thermoelectric tensors Seebeck $\beta$ and Peltier $\pi$. The Seebeck effect relates a temperature gradient $\nabla T$ with the appearance of an electric field [$E_i = \beta_{ij}\nabla_jT$], and the Peltier effect connects an electric field with a heat flux density ${\bf q}$ ($q_i = \pi_{ij} E_j$). For the Seebeck and Peltier effects the Onsager relations lead to $\{-1||1\}\beta_{ij}=\pi_{ji}$ and $\{-1||1\}\pi_{ij}=\beta_{ji}$ \cite{Gallego2019}. It is then interesting to take the combinations $\frac{1}{2}(\beta_{ij} + \pi_{ji})$ and $\frac{1}{2}(\beta_{ij} - \pi_{ji})$, which are invariant and anti-invariant under time reversal respectively \cite{Grimmer2017}. From these behaviors under $\{-1 || 1\}$, and following the same reasoning as in the case of $\rho^s$ and $\rho^a$, it can be deduced that $\frac{1}{2}(\beta_{ij} + \pi_{ji})$ must be a V$^2$ tensor and $\frac{1}{2}(\beta_{ij} - \pi_{ji})$ must be of $a$V$^2$ type. The former is responsible for the ordinary Seebeck effect and the latter for the so-called spontaneous Nernst effect.

Table \ref{t-transport-properties} contains a summary of the transformation properties of the spin contributions for some transport tensors. As with the tensors discussed in the previous section, in the case of those having Jahn symbols without the letter "M", the additional constraints resulting from the SpPG can be simply obtained by comparing their symmetry-adapted forms under \mpgeff~with those under the actual MPG of the structure. A table including more transport properties together with the separation of their orbital and spin parts, where applicable, is shown in the Supporting Information (Table S3).
\begin{table}[ht]
	\centering
	\caption{\label{t-transport-properties}Selected examples of the spin contributions of some transport tensors and their Jahn symbols in the context of MPGs and SpPGs. For the SpPGs the transformation law that each Jahn symbol implies are also given.}
	\begin{tabular}{|c|l|l|l|}
		\hline
		\textbf{Tensor Description} & \textbf{Defining} & \textbf{Jahn Symbol} & \textbf{Transformation} \\
		 & \textbf{Equation} & \textbf{(MPG/SpPG)} & \textbf{laws (SpPG)} \\
		\hline
		Hall effect tensor $R^s_{ijk}$ & $E_i = R_{ijk} J_j H_k$ & $ae$\vsim V/\vsim M & $RRUR^s$ (Spin)\\
		(symmetric part) &  $R^s_{ijk} = \frac{1}{2}\left(R_{ijk} + R_{jik}\right)$ & & \\
		Linear magnetoresistance &  & & \\
		\hline
		Hall effect tensor $R^a_{ijk}$ & $E_i = R_{ijk} J_j H_k$  & $e$\{V$^2$\}V/$a$\{V$^2$\}M & $\det(U) RRUR^a$ (Spin) \\
		(antisymmetric part) &  $R^a_{ijk} = \frac{1}{2}\left(R_{ijk} - R_{jik}\right)$ & & \\
		Ordinary Hall effect &  & & \\
		\hline
		Spin/orbital Hall resistivity  & $E_i={\rho_{ij}}^k{J_j}^k$ & $ae$\vsim V/\vsim M & $RRU\rho^s$ (Spin) \\
		tensor ${\rho^s_{ij}}^k$ (symmetric part) & ${\rho^s_{ij}}^k=\frac{1}{2}\left({\rho_{ij}}^k+{\rho_{ji}}^k\right)$ &  &  \\
		\hline
		Spin/orbital Hall resistivity & $E_i={\rho_{ij}}^k{J_j}^k$ & $e$\{V$^2$\}V/$a$\{V$^2$\}M & $\det(U)RRU\rho^a$ (Spin) \\
		tensor ${\rho^a_{ij}}^k$ (antisymmetric part) & ${\rho^a_{ij}}^k=\frac{1}{2}\left({\rho_{ij}}^k-{\rho_{ji}}^k\right)$ &  &  \\
		\hline
		\hline
	\end{tabular}
\end{table}
  \subsection{Constraints on transport tensors of collinear and coplanar magnetic structures}
\label{s-transport-constraints}
Similarly to equilibrium tensors, the minimal intrinsic spin-only subgroup of collinear and coplanar structures can impose important restrictions on tensors describing transport properties. A compilation of these restrictions for the spin contributions of some properties is shown in Table \ref{t-transport-constraints}. In the Supporting Information we show further transport properties and separate the constraints for the orbital and spin parts where relevant (Table S4). In some cases, the constraints are very important. For example, the antisymmetric part of the resistivity is forbidden in collinear and coplanar structures, and therefore, the anomalous Hall effect can only be non-relativistic in non-coplanar magnetic structures \cite{Taguchi2001,Nagaosa2010}. The same happens for the spontaneous Nernst and Ettingshausen effects. Similarly, the spin Hall resistivity tensor is highly restricted in collinear and coplanar materials, with the antisymmetric part of the tensor totally vanishing in the case of collinear structures \cite{Zhang2018}. In contrast, the orbital contribution of the antisymmetric part of the Hall effect tensor (orbital part of the ordinary Hall effect, see Table S4 in the Supporting Information) is not restricted by the collinearity or coplanarity, and in fact that property can exist in materials of any symmetry.
\begin{table}[ht]
	\centering
	\caption{\label{t-transport-constraints}Constraints imposed by collinearity and coplanarity on the spin contributions of some tensors for transport phenomena assuming SpPG symmetry. The $z$ direction is chosen as in Table \ref{t-equilibrium-properties} to define the orientation of the spins or the spin planes.}
	\begin{tabular}{|c|l|l|}
		\hline
		\textbf{Tensor} & \textbf{Collinear Structure} & \textbf{Coplanar Structure} \\
		\hline
		Hall effect tensor $R^s_{ijk}$  & $R^s_{ij1} = R^s_{ij2} = 0$, & $R^s_{ij1}$, $R^s_{ij2}$ no restriction,\\
		(symmetric part) (spin contribution) &  $R^s_{ij3}$ no restriction & $R^s_{ij3}=0$ \\
		Linear magnetoresistance &   &  \\
		\hline
		Hall effect tensor $R^a_{ijk}$ & $R^a = 0$ & $R^a_{ij1} = R^a_{ij2} = 0$, \\
		(antisymmetric part) (spin contribution) &  & $R^a_{ij3}$ no restriction \\
		Ordinary Hall effect &  &  \\
		\hline
		Spin Hall resistivity tensor ${\rho^s_{ij}}^k$& ${\rho^s}^1 = {\rho^s}^2 = 0$, & ${\rho^s}^1$, ${\rho^s}^2$ no restriction, \\
		(symmetric part) & ${\rho^s}^3$ no restriction & ${\rho^s}^3 = 0$ \\
		\hline
		Spin Hall resistivity tensor ${\rho^a_{ij}}^k$ & ${\rho^a}^1 = {\rho^a}^2 = {\rho^a}^3 = 0$ & ${\rho^a}^1 = {\rho^a}^2 = 0$, \\
		(antisymmetric part) &  &  ${\rho^a}^3$ no restriction \\
		\hline
	\end{tabular}
\end{table}  \subsection{Optical properties}
\label{s-optical-properties}
The optical behavior of a material is based on the properties of its dielectric permittivity tensor at high frequencies $\varepsilon_{ij}$, as well as on the changes that this tensor undergoes when the material is subjected to external influences (magnetic fields, electric fields, stress...). As we have pointed out in our study of equilibrium properties, the permittivity tensor is of type \vsim~ for static electric fields. However, at optical frequencies the material response is not in equilibrium. It can be shown that Onsager's relations give rise to an expression similar to equation (\ref{e-magnetoelectric-orbital}) for the action of time reversal on the optical dielectric tensor \cite{Eremenko1992}, i.e.,
\begin{equation}
	\{-1 || 1\}\varepsilon_{ij} = \varepsilon_{ji}
\end{equation}

Following the same reasoning as for the resistivity, the separation into symmetric and antisymmetric parts, $\varepsilon= \varepsilon^s + \varepsilon^a$, even and odd for time reversal, gives rise to the following Jahn symbols: \vsim~ for $\varepsilon^s$, and $a$\{V$^2$\} for $\varepsilon^a$. The symmetric term describes the index ellipsoid and the antisymmetric part the spontaneous Faraday effect.

The variation of $\varepsilon_{ij}$ due to the space dispersion (dependence with the light wave vector ${\bf k}$), the application of an electric field and the application of a magnetic field can be written respectively as
\begin{eqnarray}
	\label{e-susceptibility-k}
	\varepsilon_{ij}({\bf k}) = \varepsilon_{ij}(0) + i \gamma_{ij\ell} k_{\ell} + \gamma^{(2)}_{ij\ell m} k_{\ell} k_m + \cdots,\\
	\label{e-susceptibility-E}
	\varepsilon_{ij}({\bf E}) = \varepsilon_{ij}(0) + r_{ijk} E_k + R_{ijk\ell} E_k E_{\ell} + \cdots,\\
	\label{e-susceptibility-H}
	\varepsilon_{ij}({\bf H}) = \varepsilon_{ij}(0) + i z_{ijk} H_k + R_{ijk\ell} H_k H_{\ell} + \cdots
\end{eqnarray}
Again, if we separate $\varepsilon_{ij}$ into symmetric and antisymmetric parts, and take into account the properties of transformation of ${\bf E}$, ${\bf H}$, and ${\bf k}$ (the latter being a rank 1 tensor that changes sign both under inversion $\{1 || \overline{1}\}$ and under time reversal $\{-1 || 1\}$), we can easily deduce the Jahn symbols of the various tensors involved. A summary of some of the effects up to rank 3 is given in Table S5 of the Supporting Information. Table \ref{t-faraday} shows just the case of the spin contribution to the Faraday effect tensors (symmetric and antisymmetric parts), where the Jahn symbols for SpPGs are different from those for MPGs.
\begin{table}[ht]
	\centering
	\caption{\label{t-faraday}Spin contribution to the Faraday effect tensors with their Jahn symbols in the context of MPGs and SpPGs, and their transformation laws under an SpPG operation.}
	\begin{tabular}{|c|l|l|l|}
		\hline
		\textbf{Tensor Description} & \textbf{Defining} & \textbf{Jahn Symbol} & \textbf{Transformation} \\
		 & \textbf{Equation} & \textbf{(MPG/SpPG)} & \textbf{laws (SpPG)} \\
		\hline
		Faraday effect tensor $z^s_{ijk}$ & $\varepsilon_{ij}({\bf H}) = \varepsilon_{ij}(0) + i z_{ijk} H_k$  & $ae$\vsim V/\vsim M & $RRUz^s$ (Spin)\\
		 (symmetric part) & $z^s_{ijk} = \frac{1}{2}\left(z_{ijk} + z_{jik}\right)$ &  & \\
		Magnetooptic Kerr effect (MOKE) &  &  &  \\
		\hline
		Faraday effect tensor $z^a_{ijk}$ & $\varepsilon_{ij}({\bf H}) = \varepsilon_{ij}(0) + i z_{ijk} H_k$ &  $e$\{V$^2$\}V/$a$\{V$^2$\}M & $\det(U)RRUz^a$ (Spin)\\
		(antisymmetric part) &  $z^a_{ijk} = \frac{1}{2}\left(z_{ijk} - z_{jik}\right)$ & & \\
		Ordinary Faraday effect &  &  &  \\
		\hline
	\end{tabular}
\end{table}
 
If the medium is non-dissipative it can be shown that $\varepsilon_{ij}$ must be Hermitian \cite{Landau1960}, i.e., $\varepsilon_{ij} = \varepsilon_{ji}^{*}$. If this situation arises, it can be easily shown that the symmetric and antisymmetric parts of the various tensors must be real or pure imaginary. So, for example, for the Pockels tensor $r$ defined in equation (\ref{e-susceptibility-E}), the symmetric part  $r^s$ $\left[r^s_{ijk}=\frac{1}{2}(r_{ijk}+r_{jik})\right]$ is a real tensor and the antisymmetric part $r^a$ $\left[r^a_{ijk}=\frac{1}{2}(r_{ijk}-r_{jik})\right]$ is pure imaginary. The presence of $i$ in equations (\ref{e-susceptibility-k}) and (\ref{e-susceptibility-H}) makes the antisymmetric part of $\gamma_{ij\ell}$ and $z_{ijk}$ (natural optical activity and ordinary Faraday effect) real.
 \subsection{Constraints on optical tensors of collinear and coplanar magnetic structures }
\label{s-optical-constraints}
As in the preceding cases, collinearity and coplanarity also impose restrictions on tensors for optical properties, as is shown in Table S6 of the Supporting Information. Since some optical tensors share the same Jahn symbol with some of the transport tensors listed in Tables \ref{t-transport-properties} and S3, their constraints can also be deduced from those tables. For example, the spontaneous Faraday effect, the spin contribution of the ordinary Faraday effect, and the spin contribution of the magnetooptic Kerr effect tensors have the same shape as the antisymmetric part of the resistivity, the antisymmetric part of the spin Hall resistivity, and the symmetric part of the spin Hall tensors, respectively. Consequently, \Psoin~already restricts greatly the form of these tensors both in collinear and coplanar structures. Other properties that can readily be shown to vanish for collinear and coplanar structures under SpSG symmetry are the spontaneous gyrotropic birefringence and the antisymmetric part of Pockels effect (Table S6). Table \ref{t-faraday-constraints} shows as an example the restrictions for the spin contributions to the Faraday tensors.
\begin{table}[ht]
	\centering
	\caption{\label{t-faraday-constraints}Constraints imposed by collinearity and coplanarity on the spin contributions to the Faraday tensors assuming SpPG symmetry.}
	\begin{tabular}{|c|l|l|}
		\hline
		\textbf{Tensor} & \textbf{Collinear Structure} & \textbf{Coplanar Structure} \\
		\hline
		Faraday effect tensor $z^s_{ijk}$ & $z^s_{ij1} = z^s_{ij2} = 0$, & $z^s_{ij1}, z^s_{ij2}$ no restriction, \\
		(symmetric part) (spin contribution) & $z^s_{ij3}$ no restriction & $z^s_{ij3} = 0$ \\
		Magnetooptic Kerr effect (MOKE) &  & \\
		\hline
		Faraday effect tensor $z^a_{ijk}$ & $z^a = 0$ & $z^a_{ij1} = z^a_{ij2} = 0$, \\
		(antisymmetric part) (spin contribution) &  & $z^a_{ij3}$ no restriction \\
		Ordinary Faraday effect &  &  \\
		\hline
	\end{tabular}
\end{table} 

In the Supporting Information we complete our study of crystal tensors giving an account of the transformation properties (Section S2) and constraints (Section S3) given by the SpPGs on some nonlinear optical (NLO) properties. The main conclusion is that such tensors can be studied on the basis of the \mpgeff~exclusively. \section{Examples}
\label{s-examples}
In the following we will present several examples of experimental magnetic structures with non-coplanar, coplanar, and collinear ordering for which we will obtain the symmetry adapted tensor forms for some selected properties. All the examples have been retrieved from the MAGNDATA database of the Bilbao Crystallographic Server \cite{Gallego2016}. 

We will introduce examples of the two types of magnetic structures that can be distinguished regarding the relation of their MSG and SpSG, which were discussed in Section \ref{s-relation-spin-magnetic}. These two types are on the one hand, the structures with a minimal SpSG, where the SpPG only differs from the MPG by the inclusion of the intrinsic spin-only subgroup \Psoin~(if collinear or coplanar), and the remaining ones, where the MPG is a strict subgroup of the SpPG, with the SpPG having additional space operations and/or non-trivial spin-only operations $\{U||1\}$. As explained in Section \ref{s-relation-spin-magnetic}, in order to determine the relation between the MSG of a magnetic structure and its SpSG, the SpSG must be described choosing the orientation of the spin operations with respect to the lattice, consistently with the observed structure. \subsection{ Structures with a minimal SpSG }
\label{s-minimalSpSG}
As has been pointed out in Section \ref{s-relation-spin-magnetic}, a majority of the reported magnetic structures have a minimal possible SpSG with respect to their MSG, where the family group F of the MSG is equal to the space group G$_0$ of the space operations $\{R|{\bf t}\}$ of the SpSG. Under these conditions, the MPG and the SpPG have the same set of lattice operations $R$, and the SpPG \Ps~can be written as $\textrm{P}_\textrm{S}=\textrm{P}_\textrm{\footnotesize{M}} \times P_{\textrm{\footnotesize{SOintr}}}$, where \Pm~is the MPG of the structure and \Psoin~the corresponding intrinsic spin-only point group.

This has interesting consequences when it comes to obtaining the tensor reductions induced by the SpPG. Starting from the well-known tensor forms under the MPG symmetry (obtained for example using the MTENSOR program \cite{Gallego2019}), the constraints due to the SpPG can be found by simply adding, in the case of collinear or coplanar structures, those given by \Psoin, which we have tabulated in previous sections. In the case of non-coplanar structures the SpPG and the MPG coincide and no additional SpPG constraint exists.

We will now examine some examples of materials that illustrate the points made above.  \subsubsection{Collinear DyB$_4$ (entry 0.22 in MAGNDATA)}
\label{s-DyB4}
\begin{figure}[ht]
	\centering
	\includegraphics[width=0.5\textwidth]{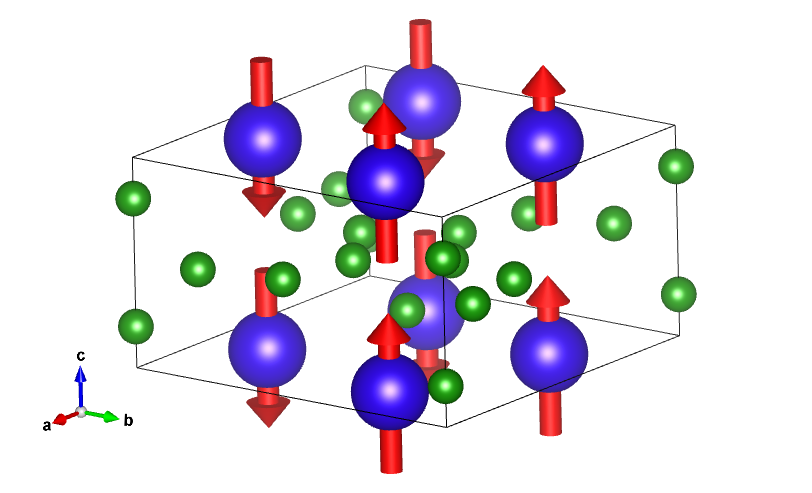}
	\caption{\label{f-DyB4}Magnetic structure of DyB$_4$ below 21K. Dy and B atoms are represented by blue and green spheres respectively.}
\end{figure}

DyB$_4$ has space group $Pbam$ (No. 55) in its paramagnetic phase and below 21K exhibits a collinear magnetic structure \cite{Will1979}, with propagation vector ${\bf k}=0$ and MSG $Pb'am$ (OG No. 55.3.433), and therefore its MPG is $m'mm$. The spins are oriented along $c$. A scheme of the structure is displayed in Fig. \ref{f-DyB4}. As the MSG keeps all the space operations of the parent space group $Pbam$, then the corresponding SpSG is minimal, with no additional space operation. This SpSG has been identified as $P\,^{-1}b\,^1a\,^1m\,^{\infty m}1$ (No. 26.55.1.1) in the so-called international notation \cite{Chen2024}, but one should take care that in this SpSG notation the $x$ and $y$ axes of the lattice have been interchanged with respect to the basis of the MSG $Pb'am$. This means that keeping the same basis as in the MSG, the nontrivial SpPG can be denoted as $^1m\,^{-1}m\,^1m$, which is generated by the operations: $\{1 || m_x\},\{-1 || m_y\}$ and $\{1 || m_z\}$. We can then write
\begin{equation}
	\label{e-SpSG-DyB4}
^1m\,^{-1}m\,^1m\,^{\infty_zm}1=m'mm\times \,^{\infty_zm}1
\end{equation}
 We will use equation (\ref{e-SpSG-DyB4}) to deduce, as an example, the constraints of the magnetoelectric tensor (inverse effect) under the SpPG (see Table \ref{t-equilibrium-properties}). For the MPG $m'mm$ we have

\begin{displaymath}
\alpha^T(m'mm)=\left(\begin{array}{rrr}0&0&0\\0&0&\alpha^T_{23}\\0&\alpha^T_{32}&0\end{array}\right)
\end{displaymath}

as can be easily checked. But the additional spin-only group $^{\infty_zm}1$ in the SpPG cancels out the elements of the first two rows (see Table \ref{t-equiligrium-constraints}). Therefore, the final tensor form under the SpPG symmetry is simply

\begin{equation}
	\label{e-magnetoelectric-DyB3-SpPG}
	\alpha^T=\left(\begin{array}{rrr}0&0&0\\0&0&0\\0&\alpha^T_{32}&0\end{array}\right)
\end{equation}

It is interesting to analyze the same case but assuming now that the spins are aligned along $a$ or $b$. The counterparts of equation (\ref{e-SpSG-DyB4}) are
\begin{displaymath}
	^1m\,^{-1}m\,^1m\,^{\infty_xm}1=mmm'\times \,^{\infty_xm}1
\end{displaymath}
and
\begin{displaymath}
	^1m\,^{-1}m\,^1m\,^{\infty_ym}1=m'm'm'\times \,^{\infty_ym}1
\end{displaymath}
In the first case, the MPG $mmm'$ gives a tensor

\begin{displaymath}
	\alpha^T(mmm')=\left(\begin{array}{rrr}0&\alpha^T_{12}&0\\\alpha^T_{21}&0&0\\0&0&0\end{array}\right)
\end{displaymath}
and in the second

\begin{displaymath}
	\alpha^T(m'm'm')=\left(\begin{array}{rrr}\alpha^T_{11}&0&0\\0&\alpha^T_{22}&0\\0&0&\alpha^T_{33}\end{array}\right)
\end{displaymath}

For these orientations, $^{\infty_xm}1$ eliminates the second and third rows of $\alpha^T$, while $^{\infty_ym}1$ does the same with the first and third rows. Then we have under the SpPG
\begin{equation}
	\label{e-magnetoelectric-DyB3-SpPG-x}
	\alpha^T=\left(\begin{array}{rrr}0&\alpha^T_{12}&0\\0&0&0\\0&0&0\end{array}\right)
\end{equation}
and
\begin{equation}
	\label{e-magnetoelectric-DyB3-SpPG-y}
	\alpha^T=\left(\begin{array}{rrr}0&0&0\\0&\alpha^T_{22}&0\\0&0&0\end{array}\right)
\end{equation}
respectively.

These three results are easily interpretable. The three tensor forms for the three spin directions, equations (\ref{e-magnetoelectric-DyB3-SpPG})-(\ref{e-magnetoelectric-DyB3-SpPG-y}), correspond to the same physical effect under the SpPG. They simply indicate that the electric induced magnetization can only take place along the spin directions, which without SOC would be arbitrary. In contrast, independently of the direction of the spins, the electric field must be applied along a specific crystal direction, namely the $y$-axis, which is the direction perpendicular to the unique mirror plane with $U=-1$ in the non trivial SpPG. As can be seen with this example, physically equivalent tensor reductions under the same SpPG, for different orientations of the spins, can be derived starting from tensor forms under different MPGs. 

Equations (\ref{e-magnetoelectric-DyB3-SpPG-x}) and (\ref{e-magnetoelectric-DyB3-SpPG-y}) could have been deduced from equation (\ref{e-magnetoelectric-DyB3-SpPG}) by using equation (\ref{e-magnetoelectric-rotated}), which relates the $\alpha^T$ tensors in structures differing in their spin orientations. Using this procedure we easily obtain that the only surviving coefficient in equations (\ref{e-magnetoelectric-DyB3-SpPG})-(\ref{e-magnetoelectric-DyB3-SpPG-y}) must have numerically the same value.

In the description proposed at the end of Section \ref{s-equilibrium-properties}, the magnetoelectric tensor of this example, when described with separate spin and lattice systems, would have only a single coefficient, $\alpha^T_{3'2}$, similar to equation (\ref{e-magnetoelectric-DyB3-SpPG}). This means that an electric field along the $y$ lattice direction induces a magnetization along the spin direction $z’$, whatever this may be. Equations (\ref{e-magnetoelectric-DyB3-SpPG})-(\ref{e-magnetoelectric-DyB3-SpPG-y}) are particular cases of this more general rule. \subsubsection{Collinear MnF$_2$ (entry 0.15 in MAGNDATA)}
	\label{s-MnF2}
\begin{figure}[ht]
	\centering
	\includegraphics[width=0.5\textwidth]{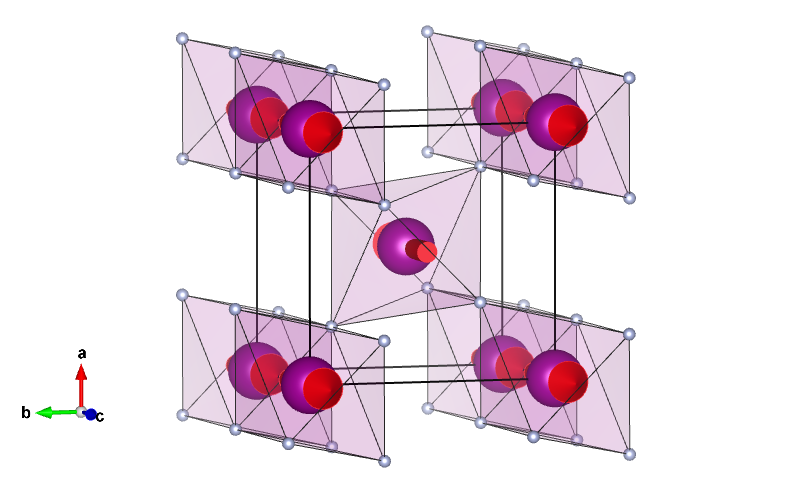}
	\caption{\label{f-MnF2}Magnetic structure of MnF$_2$ showing the spins of the Mn atoms (violet spheres). F atoms are represented by small gray spheres.}
\end{figure}

	MnF$_2$ has space group $P4_2/mnm$ (No. 136) in the paramagnetic parent phase. Upon cooling it undergoes a transition to a collinear magnetic phase with propagation vector $\mathbf{k} = 0$ \cite{Yamani2010}. The structure of the magnetic phase is shown in Fig. \ref{f-MnF2}. The spins are parallel to [001], with MSG $P4_2'/mnm'$ (OG No. 136.5.1156). Here again the MSG keeps all space operations of the parent space group $P4_2/mnm$, and therefore, the corresponding SpSG is necessarily minimal. This SpSG is $P\,^{-1}4_2/\,^1m\,^{-1}n\,^1m\,^{\infty m}1$ (No. 65.136.1.1) \cite{Chen2024}. The corresponding SpPG, $^{-1}4/\,^1m\,^{-1}m\,^1m\,^{\infty m}1$, generated by the operations, 
	\begin{displaymath}
		\{-1||4_z\}, \{1||m_z\}, \{1||m_{1\bar{1}0}\}, \{\infty_z||1\}, \{m_x||1\},
	\end{displaymath}
	can then be related with the MPG in the form
	\begin{displaymath}
	^{-1}4/\,^1m\,^{-1}m\,^1m\,^{\infty_z m}1=4'/mmm'\times \,^{\infty_z m}1
\end{displaymath}
 
	The tensor constraints according to the SpPG will then be those of the MPG plus those due to the collinearity spin-only group $^{\infty_z m}1$.
	
	We can take as an example the piezomagnetic tensor $\Lambda_{ijk}$ (see Table \ref{t-equilibrium-properties}), which has been recently considered in connection with a discussion about the altermagnetism of this material \cite{Bhowal2024,Radaelli2024}. The results are obtained straightforwardly using the \texttt{MTENSOR} program and Table \ref{t-equiligrium-constraints}.
	
	The constraints under the MPG give
	\begin{equation}
		\label{e-piezomagnetic}
		\Lambda=\left(\begin{array}{cccccc}0&0&0&\Lambda_{14}&0&0\\0&0&0&0&\Lambda_{14}&0\\0&0&0&0&0&\Lambda_{36}\\\end{array}\right)
	\end{equation}
	where the usual Voigt index contraction has been used for the last two indices of $\Lambda_{ijk}$. Adding the restrictions of Table \ref{t-equiligrium-constraints}, the only coefficient that survives is simply $\Lambda_{36}$. This means that the magnetization induced by stress is only along the spin direction, and it can be induced only upon application of a $\sigma_{12}$ ($=\sigma_6$) shear stress. Therefore, this is the non-relativistic piezomagnetic effect, which the system is expected to have even if spin-orbit coupling (SOC) is negligible.
	
	In contrast with the example of Section \ref{s-DyB4}, in this case if we consider any other hypothetical spin direction for the collinear spin arrangement, the resulting MPG will lose some space operations, and therefore the SpSG will not be minimal with respect to the new MPG. Therefore, the simple method to derive the SpSG-adapted form of the tensors employed above is not possible for any other spin direction. But from equation (\ref{e-piezomagnetic}) we can infer how it would be the SOC-free piezomagnetic effect in any case. Taking into account that $\Lambda$ is of type M\vsim, and following a procedure similar to the one carried out for the magnetoelectric tensor in the previous example, we easily arrive at
	\begin{equation}
		\label{e-piezomagnetic-transf}
		\Lambda_{ijk}\,' = P_{i\ell} \Lambda_{\ell jk}
	\end{equation}
	where $\Lambda'_{ijk}$ is the piezomagnetic tensor of the new structure and $P_{i\ell}$ is the rotation matrix relating both spin orientations. In this case, equation (\ref{e-piezomagnetic-transf}) leaves as non-null elements only $\Lambda_{i6}\,'=P_{i3}\Lambda_{36}$ ($i = 1,2,3$). Thus, in the SOC-free limit the induced magnetization is always along the spin direction, whatever this is, but the applied stress must be a shear $\sigma_6$ on the crystal basal plane. In the description using separate spin and lattice reference systems (end of Section \ref{s-equilibrium-properties}) we would have here a tensor with just a single coefficient, $\Lambda_{3'6}$, meaning that a stress $\sigma_6$ induces a magnetization along the spin direction, this being arbitrary.
 \subsubsection{Coplanar CoSO$_4$ (entry 1.519 in MAGNDATA).}
\label{s-CoSO4}
\begin{figure}[ht]
	\centering
	\includegraphics[width=0.5\textwidth]{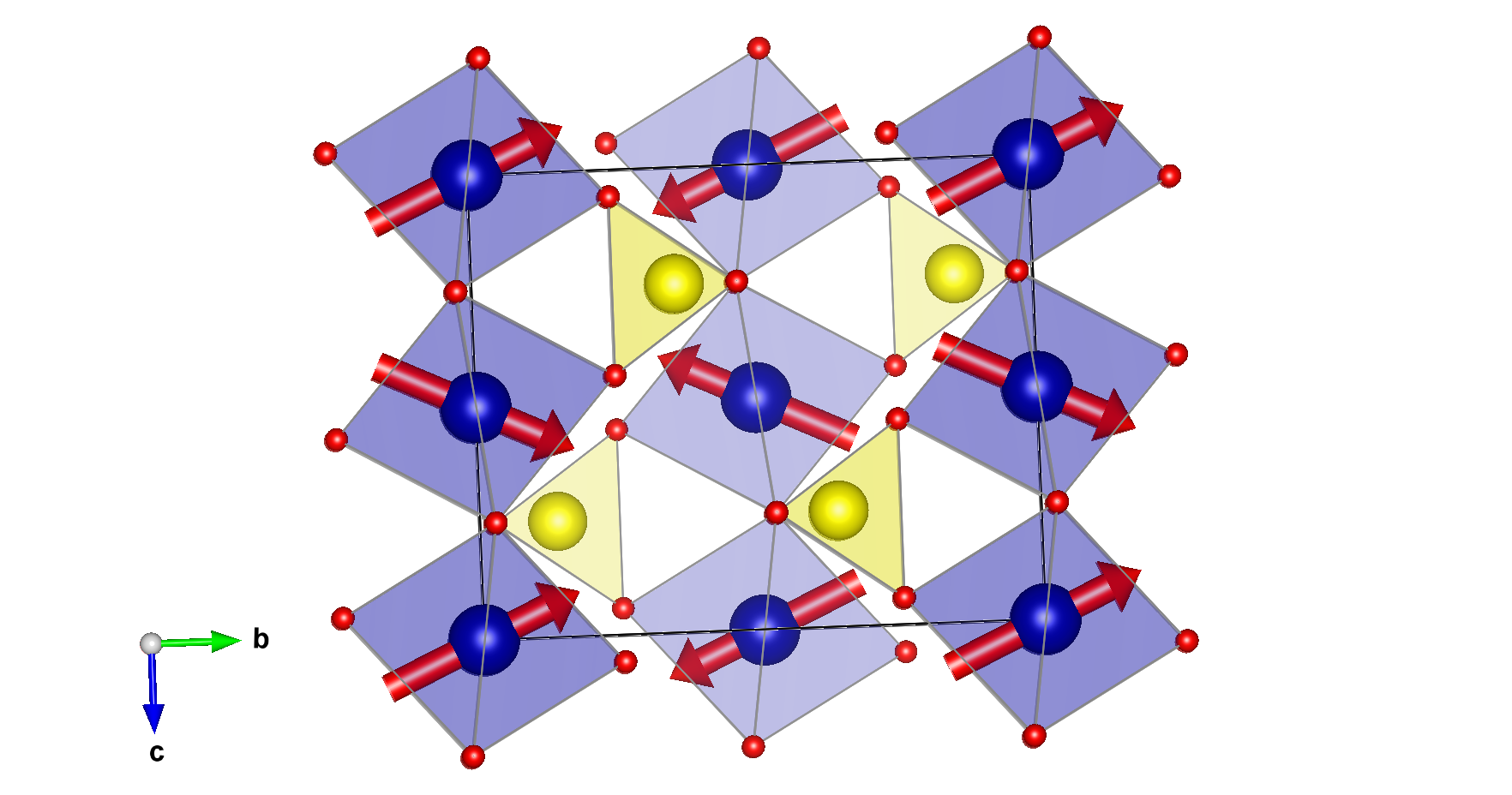}
	\caption{\label{f-CoSO4}Magnetic structure of CoSO$_4$ below 15.5K showing the spins of the Co atoms (blue spheres). The O and S atoms are represented by red and yellow spheres respectively.}
\end{figure}

	CoSO$_4$ has a paramagnetic phase with space group $Cmcm$ (No. 63), and a magnetic phase below 15.5~K with propagation vector $\mathbf{k} = (1,0,0)$ \cite{Frazer1962}. The material is coplanar, being $m_x$ the spin-only mirror plane (see Fig. \ref{f-CoSO4}). Its MSG is $P_Cbcn$ in the BNS notation, with OG numerical index 63.16.52. The non-trivial SpSG is 10.63.2.1 \cite{Chen2024}. Also in this case, despite the non-zero propagation vector, which implies the breaking of the body-centering lattice translation, all operations of the parent space group are maintained in the MSG. The lost centering translation is kept in the MSG as an antitranslation, i.e., a translation combined with time reversal. Thus, the MPG of the structure is $mmm.1'$, and the SpPG is necessarily minimal with respect to it. The SpPG can be written as the direct product of the MPG and the coplanar spin-only group: $mmm.1' \times \,^{m_x}1$. The SpPG tensor constraints can be derived, as in previous examples, by adding to the constraints of the MPG those of the $\{m_x || 1\}$ plane of $^{m_x}1$.

	Let us consider the spin Hall resistivity tensor as an example. The MPG restricts the antisymmetric part of that tensor to the form
	
	\begin{equation}
		\label{e-hall-resistivity}
		{\rho^a}^1 = \left(\begin{array}{ccc}
		0&0&0\\0&0&{\rho_{23}}^1\\0&-{\rho_{23}}^1&0
		\end{array}\right),
		{\rho^a}^2 = \left(\begin{array}{ccc}
		0&0&{\rho_{13}}^2\\0&0&0\\-{\rho_{13}}^2&0&0
		\end{array}\right),
		{\rho^a}^3 = \left(\begin{array}{ccc}
		0&{\rho_{12}}^3&0\\-{\rho_{12}}^3&0&0\\0&0&0
		\end{array}\right)
	\end{equation}
	
	We can now add the additional SpPG constraints due to the coplanarity. According to Table \ref{t-transport-constraints}, the SpPG only allows a non-zero ${\rho^a}^1$ (note that the plane in \Pso~is $m_x$ instead of $m_z$) while it forces ${\rho^a}^2$ and ${\rho^a}^3$ to be null. If the tensor is expressed using separate spin and lattice reference frames, the only surviving term is ${\rho_{23}}^{3'}=-{\rho_{32}}^{3'}$, where $z'$ is the direction perpendicular to the spins plane, whatever its orientation with respect to the lattice.
	
	In this case the symmetric part of the spin resistivity is already zero under the MPG since this group contains the time reversal operation and this part of the tensor is time-odd when considered for the MPG operations (see Table \ref{t-transport-properties}).  \subsection{Structures with a non-minimal SpSG}
\label{s-nonminimalSpSG}
In the examples that we will consider in this section there are non-trivial differences between the space operations in the MPG and SpPG of the structures and/or the spin-only group \Pso~in the SpPG is larger than \Psoin. In this case \Ps~cannot be written as a product $\textrm{P}_\textrm{\footnotesize{M}} \times \textrm{P}_{\textrm{\footnotesize{SOintr}}}$.  We will take two materials (and another two in sections S5 and S6 of the Supporting Information) with different spin configurations, non-coplanar, coplanar and collinear, and we will review for them a certain set of selected properties, where we will compare the symmetry-adapted form of the corresponding tensors for the MPG and SpPG symmetries. \subsubsection{Coplanar Mn$_3$Ge (entry 0.377 in MAGNDATA)}
	\label{s-Mn3Ge}
\begin{figure}[ht]
	\centering
	\includegraphics[width=0.5\textwidth]{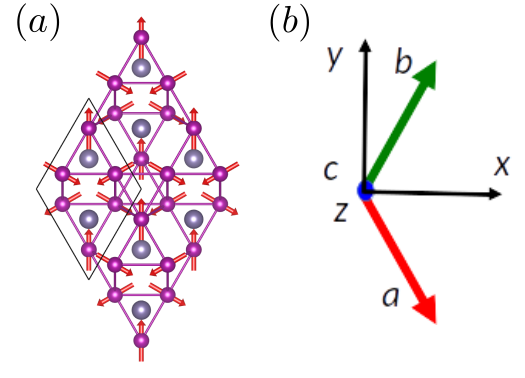}
	\caption{\label{f-Mn3Ge}(a) Magnetic structure of Mn$_3$Ge, showing the spins of the Mn atoms. (b) Relationship between the hexagonal unit cell vectors ${\bf a},{\bf b},{\bf c}$ and the orthorhombic $xyz$ directions of the basis unit vectors used to express the material tensors in the standard setting of its MSG, $Cm'cm'$.}
\end{figure}

	The paramagnetic phase of Mn$_3$Ge is hexagonal with space group $P6_3/mmc$ (No. 194). Below 380K the material undergoes a transition to a coplanar magnetic structure \cite{Soh2020}. The plane of spins is perpendicular to the hexagonal axis (see Fig. \ref{f-Mn3Ge}(a)), and the MSG of the structure is $Cm'cm'$ (OG No. 63.8.58). The corresponding MPG is $m_x'm_ym_z'$, where the $x,y,z$ axes are associated with the orthorhombic unit cell $({\bf a}+{\bf b},-{\bf a}+{\bf b},{\bf c})$ of the MSG standard unit cell. The relation of these orthorhombic axes with the crystallographic hexagonal ${\bf a},{\bf b},{\bf c}$ unit cell vectors is depicted in Fig. \ref{f-Mn3Ge}(b).

	The SpSG of this structure is a coplanar group with a nontrivial SpSG having the numerical index 11.194.1.2 \cite{Chen2024}. The SpPG is generated by the following operations (not a minimal set, to facilitate comparison with the MPG):
	\begin{displaymath}
		\textrm{SpPG:}\{m_z||1\},\{1||m_z\}, \{m_x||m_x\}, \{1||\overline{1}\}, \{3_z||6_z\},
	\end{displaymath}
	while the generators of the orthorhombic MPG are
	\begin{displaymath}
		\textrm{MPG:} \{m_x||m_x\}, \{1||\overline{1}\}, \{m_z||m_z\},
	\end{displaymath}
	where we have used the same reference system of orthorhombic axes $x,y,z$ for both the spin and the space operations. The SpPG contains the MPG, as it should, and adds two additional generators: the three-fold/six-fold rotation and the spin-only mirror plane. The requirement of tensor invariance for these two operations is sufficient to derive the additional constraints on the tensors under the SpPG. The \mpgeff~corresponding to the above SpPG, to be considered for orbital contributions, is $6/mmm.1'$. As in all coplanar and collinear structures, it is a gray magnetic group, which forbids any orbital contribution to any time-odd tensor.
	
	Table~8 gathers a few examples of tensors, showing the difference in their symmetry-adapted forms under SpPG and MPG symmetries. Some comments on the results are in order. The SpPG does not allow the existence of spontaneous magnetization, unlike the MPG. This implies that the allowed ferromagnetism of this material, which is observed macroscopically as a weak feature \cite{Soh2020}, has the SOC as the ultimate cause. Remarkably, the anomalous Hall effect, described by the antisymmetric terms of the resistivity, $\rho_{13}=-\rho_{31}$, has been reported to be "giant" \cite{Kiyohara2016}, though it should also be a SOC effect, since it is allowed by the MPG and forbidden by the SpPG.

	\begin{table}[h!]
	\centering
	\caption{\label{t-Mn3Ge}Comparison of symmetry-adapted tensor forms of some selected tensor properties in the magnetic phase of Mn$_3$Ge according to the magnetic and spin point groups.}
	\begin{footnotesize}
		\begin{tabular}{|c|c|c|}
			\hline
			\textbf{Tensor}                  & MPG  &  SpPG   \\
			\textbf{Property}                  & &    \\
			\hline
			Magnetization                             & $(0, M_2, 0)$           &  $(0,0,0)$                      \\
			\hline
			Magnetic         &  \multirow{4}{*}{$\left(\begin{array}{ccc}\chi_{11}&0&0\\0&\chi_{22}&0\\0&0&\chi_{33}\end{array}\right)$}  &  \multirow{4}{*}{$\left(\begin{array}{ccc}\chi_{11}&0&0\\0&\chi_{11}&0\\0&0&\chi_{33}\end{array}\right)$}  \\
			susceptibility / &    &     \\
			Electric         &    &      \\ 
			susceptibility   &    &     \\
			\hline
			Electric         &  \multirow{3}{*}{$\left(\begin{array}{ccc}\rho_{11}&0&\rho_{13}\\0&\rho_{22}&0\\-\rho_{13}&0&\rho_{33}\end{array}\right)$}  &     
			\multirow{3}{*}{$\left(\begin{array}{ccc}\rho_{11}&0&0\\0&\rho_{11}&0\\0&0&\rho_{33}\end{array}\right)$} \\ 
			resistivity   &    &     \\
			&    &     \\
			\hline
			Spin Hall    & \multirow{3}{*}{$\left(\begin{array}{ccc}0&{\rho_{12}}^1&0\\{\rho_{12}}^1&0&0\\0&0&0\end{array}\right),
					\left(\begin{array}{ccc}{\rho_{11}}^2&0&0\\0&{\rho_{22}}^2&0\\0&0&{\rho_{33}}^2\end{array}\right),
					$}   &  \multirow{3}{*}{$\left(\begin{array}{ccc}0&{\rho_{12}}^1&0\\{\rho_{12}}^1&0&0\\0&0&0\end{array}\right),
				\left(\begin{array}{ccc}-{\rho_{12}}^1&0&0\\0&{\rho_{12}}^1&0\\0&0&0\end{array}\right),
				$}                                     \\
			resistivity    &          &                                     \\
			(symmetric    &          &                                     \\
			part)    &    \multirow{3}{*}{$\left(\begin{array}{ccc}0&0&0\\0&0&{\rho_{23}}^3\\0&{\rho_{23}}^3&0\end{array}\right)$}   &                   $\rho^3=0$                  \\
		     &          &                                     \\
		     &          &                                     \\
			\hline
			Spin Hall    & \multirow{3}{*}{$\left(\begin{array}{ccc}0&0&0\\0&0&{\rho_{23}}^1\\0&-{\rho_{23}}^1&0\end{array}\right),
					\left(\begin{array}{ccc}0&0&{\rho_{13}}^2\\0&0&0\\-{\rho_{13}}^2&0&0\end{array}\right),
					$}   &  \multirow{3}{*}{$\rho^1=\rho^2=0$}                                     \\
			resistivity    &          &                                     \\
			(antiymmetric    &          &                                     \\
			part)    &    \multirow{3}{*}{$\left(\begin{array}{ccc}0&{\rho_{12}}^3&0\\-{\rho_{12}}^3&0&0\\0&0&0\end{array}\right)$}   &       \multirow{3}{*}{$\rho^3=\left(\begin{array}{ccc}0&{\rho_{12}}^3&0\\-{\rho_{12}}^3&0&0\\0&0&\end{array}\right)
				$}          \\
			    &          &                                     \\
			    &          &                                     \\
			\hline
			Ordinary    & \multirow{4}{*}{$\left(\begin{array}{ccc}a_{11}&0&0\\0&a_{22}&0\\0&0&a_{33}\end{array}\right)$}  &  \multirow{4}{*}{$\left(\begin{array}{ccc}a_{11}&0&0\\0&a_{11}&0\\0&0&a_{33}\end{array}\right)$}  \\ 
			Seebeck     &   &     \\
			effect      &     &   \\
			$a_{ij} = \frac{1}{2}(\beta_{ij} + \pi_{ji})$  &&  \\
			\hline
			Spontaneous    & \multirow{4}{*}{$\left(\begin{array}{ccc}0&0&b_{13}\\0&0&0\\b_{31}&0&0\end{array}\right)$}  &  $b=0$  \\ 
			Nernst     &   &     \\
			effect      &     &   \\
			$b_{ij} = \frac{1}{2}(\beta_{ij} - \pi_{ji})$  &&  \\
			\hline
\end{tabular}
	\end{footnotesize}
\end{table}
 
	The electric and magnetic susceptibilities change from being diagonal in the MPG with 3 independent terms to having 2 of them equal in the SpPG, keeping the axial symmetry of the parent phase. A similar case happens with the ordinary Seebeck effect and the symmetric part of the electric resistivity, with a single additional constraint, $\rho_{22}=\rho_{11}$, in the SpPG. The spin Hall resistivity ${\rho_{ij}}^k$ (antisymmetric part in the $ij$ indices) also reduces from having 3 to only 1 independent coefficient. Note that the spin-only operation $\{m_z||1\}$, due to the coplanarity of the structure, is already sufficient to make the antisymmetric part of the spin resistivity vanish for $x$ and $y$ polarizations, ${\rho^{a}}^1={\rho^{a}}^2=0$ (see Table \ref{t-transport-constraints}). On the other hand, the symmetric part of ${\rho_{ij}}^k$ is also drastically reduced (5 independent coefficients under the MPG versus 1 coefficient under the SpPG). In particular, $\rho^3$ goes from being allowed in the MPG to being null in the SpPG, which can be attributed exclusively to the coplanar spin-only symmetry present in the SpPG.
 \subsubsection{Non-coplanar DyVO$_3$ (entry 0.106 in MAGNDATA)}
	\label{s-DyVO3}
\begin{figure}[ht]
	\centering
	\includegraphics[width=0.5\textwidth]{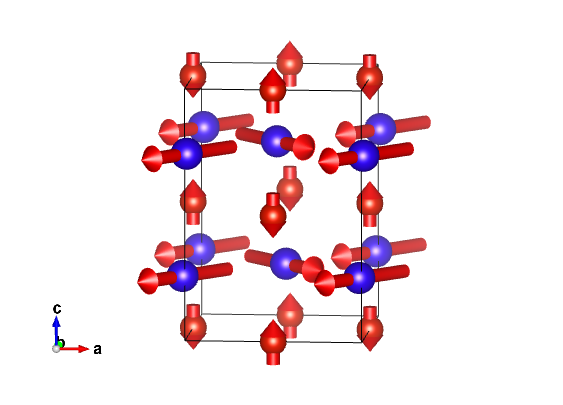}
	\caption{\label{f-DyVO3}Magnetic structure of DyVO$_3$ at 6K showing only the magnetic atoms. Blue and red spheres represent Dy and V atoms respectively.}
\end{figure}

	This material has space group $Pbnm$ (No. 62) in its paramagnetic phase. At low temperatures, both V and Dy atoms are magnetically ordered with a non-coplanar spin arrangement, which is depicted in Fig. \ref{f-DyVO3} \cite{Reehuis2011}. The MSG of this magnetic structure is $P112_1'/m'$ (OG N. 11.5.63). Being non-coplanar, the SpSG coincides with its nontrivial subgroup, which is denoted with the numerical label 2.62.1.8 in \citet{Chen2024}. Thus, the SpSG, in contrast with the MSG, keeps all the space operations of the parent space group $Pbnm$, keeping an orthorhombic symmetry, while the MSG is monoclinic. Taking as reference system the $abc$ crystallographic axes shown in Fig. \ref{f-DyVO3}, for both the spin and space operations, the corresponding SpPG can be denoted as $^{2_y}m_x\,^{m_x}m_y\,^{m_z}m_z$, which can be identified with the nontrivial SpPG with number 81 in the listing of \citet{Litvin1977}, if the labelling of the axes in the spin space is changed. As generators of this SpPG we can take: 
	\begin{displaymath}
		\{m_z||m_z\}, \{1||\overline{1}\}, \{m_x||m_y\}
	\end{displaymath}
	whereas the MPG ($2_z'/m_z'$) is generated by the first two of these three generators. Thus, the MPG is a subgroup of the SpPG, which is obtained from the former by just adding an additional generator.
 	
	With regard to the effective symmetry for the orbital contributions within the SpSG formalism, it is straightforward to derive using the SpPG generators listed above that \mpgeff$ = mm'm'$.
	
	We can now review a series of tensor properties and compare their symmetry-adapted forms according to both the MPG and SpPG.
	
	A first simple example is the spontaneous magnetization. It readily follows that the MPG allows a magnetization of the form $\mathbf{M} = (M_1, M_2, 0)$. If considered under the SpPG, as only the $U$ operations are involved in the transformations for the spin contribution to the magnetization, it can be easily seen just inspecting the mentioned above generators that under the SpPG the spin magnetization is restricted to the $y$ direction, i.e. $(0, M_2, 0)$. The magnitude of the magnetization along this direction is in fact very important, as can be seen in Fig. \ref{f-DyVO3}. In contrast, any additional spin magnetization along $x$, which is also allowed by the MPG, if present, would necessarily be a SOC effect and would break the SpPG assigned to the structure. Note however that a non-zero magnetization $M_1$ is allowed to exist without SOC, and under the same SpPG, but with the condition that it must be of orbital origin. Indeed, as \mpgeff$= mm'm'$, the orbital contribution to the magnetization must be of the form $\mathbf{M}_\textrm{\footnotesize{orb}} = (M_1, 0, 0)$, which should be added to the spin magnetization allowed along $y$.  
	
	This is an example of the problem, which was mentioned in section \ref{s-spin-groups}, that may arise in practice, when the SpSG of an experimentally determined magnetic structure is identified. Let us consider the hypothetical case of a structure like the one in this example, with negligible SOC, but with a significant orbital contribution to the atomic moments of orbital origin, resulting in a non-zero magnetization of orbital origin along $x$, as permitted by the SpSG. As the SpSG symmetry is usually determined assuming that atomic magnetic moments have only spin contributions, the observed magnetic ordering would be considered incompatible with the actual SpSG of the structure, and instead a wrong SpSG will be assigned.
	
	Another only-$U$ tensor is the spin contribution to the magnetic susceptibility $\chi^m$ (see equation (\ref{e-magnetic-susceptibility}) and Table \ref{t-equilibrium-properties}). As the SpPG maintains the orthorhombic symmetry, it is constrained to be of the form 
	\begin{equation}
		\label{e-chi-example}
		\chi^m=\left(\begin{array}{ccc}\chi^m_{11}&0&0\\0&\chi^m_{22}&0\\0&0&\chi^m_{33}\end{array}\right)
	\end{equation}
	Considering the corresponding \mpgeff, it is clear that the orbital contribution must have a similar diagonal form. Note however that, according to the rigorous definition of the SpPG, the diagonal directions $x$, $y$, and $z$ of the tensor in equation (\ref{e-chi-example}) refer only to the spin arrangement, while the diagonal axes of the orbital magnetic susceptibility are the crystallographic ones. In the SpSG formalism, the spin arrangement is considered unlocked from the lattice, and its global orientation is assumed arbitrary. Hence, if the SpPG concept is taken literally, the two diagonal tensors of spin and the orbital magnetic susceptibilities refer in general to two different systems of axes. But the clear locking between lattice and spins in a real case as this, obvious in Fig. \ref{f-DyVO3}, makes necessary that the reference axes for the spins are chosen coincident with the crystallographic ones, as we did in the description of the SpPG.
	
	As the MPG is monoclinic, the magnetic susceptibility under this lower symmetry also includes non-diagonal terms, namely the coefficient $\chi_{12}$, since the monoclinic axis is along $z$. Thus, the tensor deviation from the orthorhombic prescribed diagonal form, also valid for the paramagnetic phase, is expected to be a SOC effect.

	The same reduction as in equation (\ref{e-chi-example}) happens with other second-rank tensors like, for example, the static electric susceptibility $\chi^e$. Although the Jahn symbol of this tensor is the same for the MPG than for the SpPG (\vsim), the final form of the reduced tensor is different because of the presence of the extra space operation in the SpPG. Identical conclusions are reached for the symmetric part of the electric resistivity tensor $\rho$ or the symmetric part of the optical dielectric tensor $\varepsilon$, since their Jahn symbols are \vsim~in all cases (see Tables S3 and S5 in the Supporting Information).
	
	The antisymmetric parts of the electric resistivity tensor $\rho$ and optical dielectric tensor $\varepsilon$ have also different forms for the MPG and SpPG (see last columns in Tables S3 and S5 in the Supporting Information). The final symmetry-adapted forms of the tensors are 
	\begin{eqnarray}
		\label{e-example-resistivity}
		\rho^a & = \left(\begin{array}{ccc}0&0&\rho_{13}\\0&0&\rho_{23}\\-\rho_{13}&-\rho_{23}&0\end{array}\right),\textrm{ and }\rho^a & = \left(\begin{array}{ccc}0&0&0\\0&0&\rho_{23}\\0&-\rho_{23}&0\end{array}\right)
	\end{eqnarray}
	for the MPG and SpPG respectively, and equivalent forms for the antisymmetric part of the optical dielectric tensor. The second of equations (\ref{e-example-resistivity}) shows that even under the SpPG symmetry, and therefore with negligible SOC, the anomalous Hall effect (and the spontaneous Faraday effect) are permitted in the material. The $\rho_{23}$ component would correspond to the so-called geometric part of the Hall effect, whereas $\rho_{13}$ is a Karplus-Luttinger term, which is SOC-assisted, and should typically be proportional to $M_2$ \cite{Watanabe2024}.
	
	Further  tensor properties of this material are presented in the Supporting Information (section S4) along with other example materials: non-coplanar CaFe$_3$Ti$_4$O$_{12}$ and collinear UCr$_2$Si$_2$C (sections S5 and S6, respectively).
 \section{Conclusions}
\label{s-conclusions}
In this paper we present a general formalism for the derivation of the symmetry-adapted form of any crystal tensor property of a magnetic material considering its SpPG. We have stressed the important fact that a null SOC is required for a SpSG to be rigorously considered as a symmetry group of a magnetic structure. This means that  SpSG should be considered in most real cases as approximate symmetries. In order to compare tensor constraints under SpSG symmetry with those under the actual magnetic group of the structure, both the spin and magnetic groups must be described within a common framework, where they have a group-subgroup relation. This implies to choose a specific orientation of the spin arrangement with respect to the lattice, consistent with the observed structure. In this way, SOC-free tensor properties, permitted by the SpPG symmetry, can be systematically distinguished from those having necessarily SOC as their ultimate cause.

After reviewing the mathematical structure of SpSGs and SpPGs and their relation with ordinary MSGs and MPGs, the symmetry conditions to be satisfied by crystalline tensors under a SpPG have been analyzed. More specifically, we have carried out a systematic study of the specific action that a $\{U||R\}$ operation of a SpPG produces on various types of tensors describing macroscopic physical properties of magnetic structures. The transformation laws obtained constitute a generalization of the laws corresponding to the MPG operations, which are particular cases when $U=\pm R$. Using a generalization of the Neumann Principle to SpPGs we have found the restrictions that the SpPG symmetry imposes on 4 types of tensors, describing respectively equilibrium, transport, optical and 2nd-order NLO properties. To each tensor property we have assigned a symbol, which generalizes the Jahn symbols for the MPGs and summarizes its transformation properties under a general operation $\{U||R\}$.

We demonstrate that the spin-only symmetry, which is intrinsic in the SpPG of all collinear or coplanar magnetic structures, introduces very general constraints on the tensors when SOC-free SpPG symmetry is assumed. It is worth noting that in most practical cases (about 75\% of the reported structures), the SpSG only adds the spin-only symmetry and, therefore, the general collinear-based or coplanar-based constraints are the only extra restrictions to be added to the constraints resulting from the MPG. Finally, we illustrate the effects of the SpPG symmetries on various tensor properties for more complex SpPG-MPG relations by analyzing several examples of representative materials with non-coplanar, coplanar and collinear magnetic orderings.

A word of caution is in order regarding the way that the formalism presented in this work can be applied to an experimentally determined magnetic structure. The identification of the MSG of a given structure is a well-defined mathematical process, with no additional assumption needed, except that the structure is correct. But the determination of its SpSG, as its alternative symmetry group in the case that the SOC is null, has some ambiguities. The SpSGs of practically all commensurate magnetic structures available in the MAGNDATA database have been calculated and reported in several works \cite{Chen2024,Jiang2024,Xiao2024}. However, these SpSG identifications were done with the implicit assumption that the spin arrangement does not have any feature caused by the SOC that would falsify the calculated SpSG. This is usually quite a reasonable assumption because, except for the magnetic anisotropy that locks the global orientation of the spin arrangement with respect to the lattice, structural effects with SOC origin are usually weak. In many cases, they are not detectable by the typical neutron diffraction techniques employed in magnetic structure determination. However, this assumption sometimes fail, for instance, when the structure includes some small but significant spin canting of SOC origin. As an example, if one inspects Fig. \ref{f-CoSO4}, one may suspect that the deviation of the structure from collinearity is a local locking effect, which requires a non-zero SOC.  Thus, there are experimental structures whose assigned SpSG is a subgroup of the resulting SpSG if the SOC contribution were not considered (no canting in the example above), and the distinction between SOC-free and SOC-based tensor properties using the assigned SpSG would be wrong.  The SpSGs identified from MAGNDATA entries also assumed that the magnetic orderings have no orbital contribution or are irrelevant for the SpSG determination. We have seen above that in collinear structures the associated SpSG symmetry forbids in any case any orbital contribution to the atomic spins. There are however collinear structures with a demonstrated significant contribution to the atomic moments due to SOC effects. Hence, in such cases, ignoring the presence of the orbital contribution paradoxically allows one to assign the correct SOC-free SpSG.

Regardless of whether the calculated SpSG of an experimentally determined magnetic structure is or it is not the SOC-free symmetry group of the system, it might be tempting to consider this group as a “geometric” symmetry feature, which could be applied to derive the symmetry constraints for any property of the material. This would be, however, wrong. If the tensor constraints dictated by the identified SpSG symmetry were taken as exact, then many important observations would remain unexplained, such as the weak ferromagnetism in collinear or coplanar structures, the magnetically induced electric polarization found in many multiferroics, or the significant orbital contribution present in some collinear structures. In summary, SpSG symmetry should not be generally taken as the real symmetry of a structure, but as a good approximation, which allows one to separate, as shown in this work, those features and properties in the system which are not caused by the SOC, and therefore are especially important.

To end this paper, we would like to announce that recently we have developed a computer program (STENSOR) that, following the approach presented in this article, permits to carry out an automatic calculation of symmetry-adapted tensors under a given oriented SpPG and compare them with their form under the corresponding MPG. It is open access and has been incorporated to the Bilbao Crystallographic Server (\href{https://cryst.ehu.es/cryst/stensor.html}{cryst.ehu.es/cryst/stensor.html}). 
\acknowledgments
This work has been supported by the Government of the Basque Country (Project No. IT1458-22).

\onecolumngrid
\renewcommand{\thesection}{Appendix \arabic{section}}
\renewcommand{\thesubsection}{\arabic{section}.\arabic{subsection}}

\clearpage
\begin{center}
{\bf Supplementary material of "Crystal tensor properties of magnetic materials with and without spin-orbit coupling. Application of spin point groups as approximate symmetries."}
\end{center}

\tableofcontents

\clearpage

\addtocontents{toc}{\protect\setcounter{tocdepth}{3}}
\addtocontents{lot}{\protect\setcounter{lotdepth}{3}}
\renewcommand{\thetable}{S\arabic{table}}
\renewcommand{\thefigure}{S\arabic{figure}}
\renewcommand{\thesection}{S\arabic{section}}
\renewcommand{\thesubsection}{S\arabic{section}.\arabic{subsection}}
\setcounter{section}{0} 
\setcounter{table}{0} 
\setcounter{equation}{0} 

\section{Tables (S1-S6) of tensor properties and constraints due to spin group symmetry}
\subsection{Tensors of selected equilibrium properties}
\begin{table}[h]
	\centering
	\caption{Selection of some equilibrium properties with their Jahn symbols for the MPGs and SpPGs (added only when it is different from the symbol of the MPG), and their transformation laws under the SpPG. $\varepsilon_{ij}$ and $\sigma_{jk}$ stand for the strain and stress tensors respectively. The symbol $\varepsilon$ in the last column stands for the Levi-Civita symbol and $\overline{\alpha}$ and $b$ are the rank-2 and rank-3 tensors defined in sections 4.3 and 4.4 of the main text, respectively. In the case of MPGs, the label $e$ in the Jahn symbol indicates an axial tensor and the label $a$ a magnetic tensor, i.e., odd for time reversal. This means that the law of tensor transformation adds a change of sign for improper operations ($e$) or for operations that include time reversal ($a$). Where applicable, orbital and spin contributions have been separated.}
	\begin{tabular}{|c|l|l|l|}
		\hline
		\textbf{Tensor description}  & \textbf{Defining} & \textbf{Jahn Symbol} & \textbf{Transformation} \\
		& \textbf{Equation} & \textbf{(MPG/SpPG)} & \textbf{laws (SpPG)} \\
		\hline
		Polarization  & $P_i$ & V & $R P$ \\
		\hline
		Magnetization  & $M_i$  & $ae$V/M & $U M$ (Spin) \\
		  &   &  & $\det(U)\det(R)R M$ (Orbital) \\
		\hline
		Polar Toroidic moment  & $T_i$  & $a$V/\{MV\}  & $U R\oal$; $T=\frac{1}{2}\varepsilon\oal$ (Spin) \\
		  &   &   & $\det(U)R T$ (Orbital) \\
		\hline
		Axial Toroidic moment & $A_i$ & $e$V  & $\det(R)RA$ \\
		\hline
		Dielectric susceptibility & $P_i = \chi^e_{ij} E_j$ & \vsim  & $RR\chi^e$ \\
		tensor $\chi^e_{ij}$ &  &   &  \\
		\hline
		Magnetic susceptibility & $M_i = \chi^m_{ij} H_j$  & \vsim/\msim  & $UU\chi^m$ (Spin) \\
		tensor $\chi^m_{ij}$ &   &   & $RR\chi^m$ (Orbital) \\
		\hline
		Magnetoelectric tensor $\alpha^T_{ij}$ & $M_i = \alpha^T_{ij} E_j$ & $ae$V\sq/MV  & $UR\alpha^T$ (Spin) \\
		(inverse effect) &  &   & $\det(U)\det(R)RR\alpha^T$ (Orbital) \\
		\hline
		Electrotoroidic tensor $\theta_{ij}$ & $t_i = \theta_{ij} E_j$  & $a$V\sq/\{MV\}V  & $URRb$; $\theta =\frac{1}{2}\varepsilon b$ (Spin) \\
		(inverse effect) &   &   & $\det(U)RR b$ (Orbital) \\
		 \hline
		Piezoelectric tensor $d_{ijk}$ & $P_i = d_{ijk} \sigma_{jk}$  & V\vsim  & $RRR d$ \\
		(direct effect) &   &   &  \\
		 \hline
		Piezotoroidic tensor $\gamma_{ijk}$ & $t_i = \gamma_{ijk} \sigma_{jk}$  & $a$V\vsim/\{MV\}\vsim  & $URRR b$; $\gamma=\frac{1}{2}\varepsilon b$ (Spin)\\
		(direct effect) &   &   & $\det(U)RRR\gamma$ (Orbital) \\
		\hline
		Second order magnetoelectric & $P_i = \alpha_{ijk} H_j H_k$  & V\vsim/V\msim & $RUU \alpha$ (Spin) \\
		tensor $\alpha_{ijk}$  &   &  & $RRR \alpha$ (Orbital) \\
		(direct effect) &   &  &  \\
		\hline
		Piezomagnetic tensor $\Lambda_{ijk}$& $M_i = \Lambda_{ijk} \sigma_{jk}$  & $ae$V\vsim/M\vsim & $URR \Lambda$ (Spin) \\
		(direct effect) &   &  & $\det(U)\det(R)RRR \Lambda$ (Orbital) \\
		\hline
		Magnetostriction tensor $N_{ijk\ell}$& $\varepsilon_{ij} = N_{ijk\ell} H_kH_{\ell}$  & \vsim\vsim/\vsim\msim & $RRUUN$ (Spin) \\
		&   &  & $RRRRN$ (Orbital) \\
		\hline
	\end{tabular}
\end{table}
 \clearpage

\subsection{Constraints imposed by collinearity and coplanarity on equilibrium properties}
{\small 
	\begin{table}[h]
		\centering
		\caption{Constraints imposed by collinearity and coplanarity on some tensors of equilibrium properties. Spin and orbital contributions have been separated when applicable.}
		\begin{tabular}{|l|l|l|}
			\hline
			\textbf{Tensor}  & \textbf{Collinear structure} & \textbf{Coplanar structure} \\
			\hline
			Polarization $P_i$ & no restriction & no restriction \\
			\hline
			Magnetization $M_i$ (spin contribution)& $(0,0,M_3)$ & $(M_1,M_2,0)$ \\
			\hline
			Magnetization $M_i$ (orbital contribution)& ${\bf M} = 0$  & ${\bf M} = 0$  \\
			\hline
			Toroidic moment $T_p$  & \multirow{3}{*}{$\left(\begin{array}{ccc}0&0&0\\0&0&0\\\oal_{31}&\oal_{32}&\oal_{33}\end{array}\right)$}  & \multirow{3}{*}{$\left(\begin{array}{ccc}\oal_{11}&\oal_{12}&\oal_{13}\\\oal_{21}&\oal_{22}&\oal_{23}\\0&0&0\end{array}\right)$} \\
			(spin contribution) &   &  \\
			$T_p=\frac{1}{2}\varepsilon_{pij}\oal_{ij}$ &   & \\
			\hline
			Toroidic moment $T_i$ (orbital contribution)& ${\bf T} = 0$ & ${\bf T} = 0$ \\
			\hline
			Axial Toroidic moment $A_i$  & no restriction & no restriction \\
			\hline
			Dielectric susceptibility tensor $\chi^e_{ij}$ & no restriction  & no restriction \\
			\hline
			Magnetic susceptibility $\chi^m_{ij}$ & \multirow{3}{*}{$\left(\begin{array}{ccc}\chi^m_{11}&0&0\\0&\chi^m_{11}&0\\0&0&\chi^m_{33}\end{array}\right)$}  & \multirow{3}{*}{$\left(\begin{array}{ccc}\chi^m_{11}&\chi^m_{12}&0\\\chi^m_{12}&\chi^m_{22}&0\\0&0&\chi^m_{33}\end{array}\right)$} \\
			(spin contribution) &   &  \\
			&&\\
			\hline
			Magnetic susceptibility $\chi^m_{ij}$ (orbital contribution)& no restriction & no restriction \\
			\hline
			Magnetoelectric tensor $\alpha^T_{ij}$ & \multirow{3}{*}{$\left(\begin{array}{ccc}0&0&0\\0&0&0\\\alpha^T_{31}&\alpha^T_{32}&\alpha^T_{33}\end{array}\right)$}  & \multirow{3}{*}{$\left(\begin{array}{ccc}\alpha^T_{11}&\alpha^T_{12}&\alpha^T_{13}\\\alpha^T_{21}&\alpha^T_{22}&\alpha^T_{23}\\0&0&0\end{array}\right)$}   \\
			(spin contribution) & &   \\
			(inverse effect) & &   \\
			\hline
			Magnetoelectric tensor $\alpha^T_{ij}$ & $\alpha^T = 0$ & $\alpha^T = 0$ \\
			(orbital contribution) &  &  \\
			\hline
			Electrotoroidic tensor $\theta_{pk}$ & $b_{i1j} = b_{i2j} = 0$, & $b_{i1j}$, $b_{i2j}$ no restriction,\\
			(spin contribution) (inverse effect) & $b_{i3j}$ no restriction & $b_{i3j}=0$\\
			$\theta_{pk}=\frac{1}{2}\varepsilon_{pij}b_{ijk}$ && \\
			\hline
			Electrotoroidic tensor $\theta_{ij}$ & $\theta = 0$ & $\theta = 0$ \\
			(orbital contribution) &  &  \\
			\hline
			Piezoelectric tensor $d_{ijk}$ & no restriction  & no restriction \\
			\hline
			Piezotoroidic tensor $\gamma^T_{pk\ell}$  & $b_{i1jk} = b_{i2jk} = 0$, & $b_{i1jk}$, $b_{i2jk}$ no restriction,\\
			(spin contribution) (inverse effect) & $b_{i3jk}$ no restriction & $b_{i3jk} = 0$ \\
			$\gamma^T_{pk\ell} =\frac{1}{2}\varepsilon_{pij} b_{ijk\ell}$ &&\\
			\hline
			Piezotoroidic tensor $\gamma^T_{ijk}$ & $\gamma^T = 0$ & $\gamma^T = 0$ \\
			(orbital contribution) &  &  \\
			\hline
			Second order magnetoelectric &  \multirow{3}{*}{$\left(\begin{array}{cccccc}\alpha_{11}&\alpha_{11}&\alpha_{13}&0&0&0\\\alpha_{21}&\alpha_{21}&\alpha_{23}&0&0&0\\\alpha_{31}&\alpha_{31}&\alpha_{33}&0&0&0\end{array}\right)$} &  \multirow{3}{*}{$\left(\begin{array}{cccccc}\alpha_{11}&\alpha_{12}&\alpha_{13}&0&0&\alpha_{16}\\\alpha_{21}&\alpha_{22}&\alpha_{23}&0&0&\alpha_{26}\\\alpha_{31}&\alpha_{32}&\alpha_{33}&0&0&\alpha_{36}\end{array}\right)$}\\
			tensor $\alpha_{ijk}$ &   &  \\
			(spin contribution) (direct effect) &   &  \\
			\hline
			Second order magnetoelectric & no restriction & no restriction \\
			tensor $\alpha_{ijk}$ (orbital contribution)&  &  \\
			\hline
			Piezomagnetic tensor $\Lambda_{ijk}$ & $\Lambda_{1jk} = \Lambda_{2jk} = 0$,  &  $\Lambda_{1jk}$, $\Lambda_{2jk}$ no restriction,\\
			(spin contribution) (direct effect) & $\Lambda_{3jk}$ no restriction &  $\Lambda_{3jk}=0$ \\
			\hline
			Piezomagnetic tensor $\Lambda_{ijk}$ & $\Lambda = 0$  & $\Lambda = 0$ \\
			(orbital contribution) &   &  \\
			\hline
			Magnetostriction tensor $N_{ijk\ell}$ & \multirow{6}{*}{$\left(\begin{array}{cccccc}N_{11}&N_{11}&N_{13}&0&0&0\\N_{21}&N_{21}&N_{23}&0&0&0\\N_{31}&N_{31}&N_{33}&0&0&0\\N_{41}&N_{41}&N_{43}&0&0&0\\N_{51}&N_{51}&N_{53}&0&0&0\\N_{61}&N_{61}&N_{63}&0&0&0\end{array}\right)$}& \multirow{6}{*}{$\left(\begin{array}{cccccc}N_{11}&N_{12}&N_{13}&0&0&N_{16}\\N_{21}&N_{22}&N_{23}&0&0&N_{26}\\N_{31}&N_{32}&N_{33}&0&0&N_{36}\\N_{41}&N_{42}&N_{43}&0&0&N_{46}\\N_{51}&N_{52}&N_{53}&0&0&N_{56}\\N_{61}&N_{62}&N_{63}&0&0&N_{66}\end{array}\right)$}   \\
			(spin contribution) &  &   \\
			&  &   \\
			&  &   \\
			&  &   \\
			&  &   \\
			\hline
			Magnetostriction tensor $N_{ijk\ell}$ & no restriction  & no restriction \\
			(orbital contribution) &  &   \\
			\hline
		\end{tabular}
	\end{table}
}

\clearpage
\subsection{Tensors of selected transport properties}
\begin{table}[h]
	\centering
	\caption{Selected examples of transport tensors and their Jahn symbols in the context of MPGs and SpPGs. For the SpPGs, the transformation law that each Jahn symbol implies are also given. Some tensors have separated contributions coming from spin and orbital degrees of freedom. Seebeck and Peltier tensors $\beta$ and $\pi$ which appear in the last two rows are defined through equations $E_i = \beta_{ij}\nabla_j T$ and $q_i = \pi_{ij} E_j$. Ordinary Seebeck and Peltier tensors are transpose of each other, and the same relationship exists between spontaneous Nernst and spontaneous Ettingshausen tensors.}
	\begin{tabular}{|l|l|l|l|}
		\hline
		\textbf{Tensor description}  & \textbf{Defining} & \textbf{Jahn Symbol} & \textbf{Transformation} \\
		  & \textbf{Equation} & \textbf{(MPG/SpPG)} & \textbf{laws (SpPG)} \\
		\hline
		Resistivity tensor $\rho^s_{ij}$ & $E_i = \rho_{ij} J_j$ & \vsim & $RR\rho^s$ \\
		(symmetric part) &  $\rho^s_{ij} = \frac{1}{2}\left(\rho_{ij} + \rho_{ji}\right)$ & & \\
		Ordinary resistivity &&&\\
		\hline
		Resistivity tensor $\rho^a_{ij}$ & $E_i = \rho_{ij} J_j$  & $a$\{V$^2$\} &  $\det(U)RR\rho^a$ \\
		(antisymmetric part) & $\rho^a_{ij} = \frac{1}{2}\left(\rho_{ij} - \rho_{ji}\right)$  &  &  \\
		Spontaneous Hall effect &&&\\
		\hline
		Hall effect tensor $R^s_{ijk}$ & $E_i = R_{ijk} J_j H_k$  & $ae$\vsim V/\vsim M  & $RRU R^s$ (Spin) \\
		(symmetric part) & $R^s_{ijk} = \frac{1}{2}\left(R_{ijk} + R_{jik}\right)$  &  & $\det(U)\det(R)RRRR^s$ \\
		Linear magnetoresistance &&& (Orbital)\\
		\hline
		Hall effect tensor $R^a_{ijk}$ & $E_i = R_{ijk} J_j H_k$  & $e$\{V$^2$\}V/$a$\{V$^2$\}M  & $\det(U)RRU R^a$ (Spin) \\
(antisymmetric part) & $R^a_{ijk} = \frac{1}{2}\left(R_{ijk} - R_{jik}\right)$  &   & $\det(R)RRR R^a$ (Orbital) \\
Ordirary Hall effect &&&\\
\hline
		Spin/orbital Hall resistivity& $E_i = {\rho_{ij}}^k {J_j}^k$ & $ae$\vsim V/\vsim M & $RRU \rho^s$ (Spin) \\
		tensor ${\rho^s_{ij}}^k$ (symmetric part) & ${\rho^s_{ij}}^k = \frac{1}{2}\left({\rho_{ij}}^k + {\rho_{ji}}^k\right)$ &  & $\det(U)\det(R)RRR \rho^s$ \\
		&&& (Orbital)\\
		\hline
		Spin/orbital Hall resistivity& $E_i = {\rho_{ij}}^k {J_j}^k$ & $e$\{V$^2$\}V/$a$\{V$^2$\}M & $\det(U)RRU \rho^a$ (Spin) \\
tensor ${\rho^a_{ij}}^k$ (antisymmetric part) & ${\rho^a_{ij}}^k = \frac{1}{2}\left({\rho_{ij}}^k - {\rho_{ji}}^k\right)$ &  & $\det(R)RRR \rho^a$ (Orbital) \\
\hline
		Ordinary Seebeck effect  & $\frac{1}{2}\left(\beta_{ij} + \pi_{ji}\right)$   & V$^2$  & $RR \frac{1}{2}(\beta+\pi)$ \\
		Ordinary Peltier effect  & $\frac{1}{2}\left(\beta_{ji} + \pi_{ij}\right)$   &   &  \\
		\hline
		Spontaneous Nernst effect   & $\frac{1}{2}\left(\beta_{ij} - \pi_{ji}\right)$  & $a$V$^2$   & $\det(U)RR \frac{1}{2}(\beta-\pi)$ \\
		Spontaneous Ettingshausen effect   & $\frac{1}{2}\left(\beta_{ji} - \pi_{ij}\right)$  &   &  \\
		\hline
	\end{tabular}
\end{table}
 \clearpage
\subsection{Constraints imposed by collinearity and coplanarity on transport properties}
\begin{table}[h]
	\centering
	\caption{Constraints imposed by collinearity and coplanarity on some tensors for transport phenomena.}
	\begin{tabular}{|l|l|l|l|}
		\hline
		\textbf{Tensor} & \textbf{Collinear structure} & \textbf{Coplanar structure} \\
		\hline
		Resistivity tensor $\rho^s_{ij}$ & no restrictions & no restrictions \\
		(symmetric part) &  &  \\
		Ordinary resistivity &  &  \\
		\hline
		Resistivity tensor $\rho^a_{ij}$ & $\rho^a=0$ & $\rho^a=0$ \\
		(antisymmetric part) &  &  \\
		Anomalous Hall effect &  &  \\
		\hline
		Hall effect tensor $R^s_{ijk}$ & $R^s_{ij1}= R^s_{ij2} =0$, & $R^s_{ij1}$, $R^s_{ij2}$ no restriction, \\
		(symmetric part) (spin contribution) & $R^s_{ij3}$  no restriction& $R^s_{ij3}=0$ \\
		Linear magnetoresistance &  &  \\
		\hline
		Hall effect tensor $R^s_{ijk}$ & $R^s=0$ & $R^s=0$ \\
		(symmetric part) (orbital contribution) &  &  \\
		Linear magnetoresistance &   &  \\
		\hline
		Hall effect tensor $R^a_{ijk}$ & $R^a=0$ & $R^a_{ij1}= R^a_{ij2} =0$, \\
		(antisymmetric part) (spin contribution) &  & $R^a_{ij3} =0$ no restriction\\
		Ordinary Hall effect &  &  \\
		\hline
		Hall effect tensor $R^a_{ijk}$ & no restriction & no restriction \\
		(antisymmetric part) (orbital contribution) &  &  \\
		Ordinary Hall effect &  &  \\
		\hline
		Spin Hall resistivity tensor ${\rho^s_{ij}}^k$ & ${\rho^s}^1={\rho^s}^2=0$ &  ${\rho^s}^1$, ${\rho^s}^2$ no restriction,\\
		(symmetric part) & ${\rho^s}^3$ no restriction & ${\rho^s}^3=0$ \\
		\hline
		Spin Hall resistivity tensor ${\rho^a_{ij}}^k$ & $\rho^a=0$ & ${\rho^a}^1={\rho^a}^2=0$, \\
		(antisymmetric part) &  & ${\rho^a}^3$ no restriction \\
		\hline
		Ordinary Seebeck effect & no restriction & no restriction \\
		Ordinary Peltier effect &  &  \\
		\hline
		Spontaneous Nernst effect & $\frac{1}{2}(\beta_{ij} - \pi_{ji})=0$ & $\frac{1}{2}(\beta_{ij} - \pi_{ji})=0$ \\
		Spontaneous Ettingshausen effect & $\frac{1}{2}(\beta_{ji} - \pi_{ij})=0$ & $\frac{1}{2}(\beta_{ji} - \pi_{ij})=0$ \\
		\hline
	\end{tabular}
\end{table}
 \clearpage
\subsection{Tensors of selected optical properties}
	
\begin{table}[h]
	\centering
	\caption{Selected examples of optical properties tensors with their Jahn symbols in the context of MPGs and SpPGs, and their transformation laws under an operation $\{U\parallel R\}$.}
	\begin{tabular}{|l|l|l|l|}
		\hline
		\textbf{Tensor description} & \textbf{Defining } & \textbf{Jahn Symbol} & \textbf{Transformation} \\
		& \textbf{Equation} & \textbf{(MPG/SpPG)} & \textbf{laws (SpPG)} \\
		\hline	
			Optical dielectric tensor $\varepsilon^s_{ij}$& $D_i = \varepsilon_{ij} E_j$ & \vsim &  $RR\varepsilon^s$ \\
			(symmetric part)& $\varepsilon^s_{ij} = \frac{1}{2}(\varepsilon_{ij} + \varepsilon_{ji})$ &  &  \\
			Index ellipsoid &  &  &  \\
			\hline
			Optical dielectric tensor $\varepsilon^a_{ij}$ & $D_i = \varepsilon_{ij} E_j$ & $a$\{V$^2$\} &  $\det(U)RR\varepsilon^a$ \\
			(antisymmetric part) & $\varepsilon^a_{ij} = \frac{1}{2}(\varepsilon_{ij} - \varepsilon_{ji})$ &  &  \\
			Spontaneous Faraday effect &  &  &  \\
			\hline
			Optical activity tensor $\gamma^s_{ij\ell}$ & $\varepsilon_{ij}({\bf k}) = \varepsilon_{ij}(0) + i \gamma_{ij\ell} k_{\ell}$ & $a$\vsim V &  $\det(U)RRR\gamma^s$ \\
			(symmetric part) & $\gamma^s_{ij\ell} = \frac{1}{2}(\gamma_{ij\ell} + \gamma_{ji\ell})$ &  &  \\
			Spontaneous gyrotropic birefringence &  &  &  \\
			\hline
			Optical activity tensor $\gamma^a_{ij\ell}$ & $\varepsilon_{ij}({\bf k}) = \varepsilon_{ij}(0) + i \gamma_{ij\ell} k_{\ell}$ & \{V$^2$\}V &  $RRR\gamma^a$ \\
			(antisymmetric part) & $\gamma^a_{ij\ell} = \frac{1}{2}(\gamma_{ij\ell} - \gamma_{ji\ell})$ &  &  \\
			Natural optical activity & & & \\
			\hline
			Pockels effect tensor $r^s_{ijk}$ & $\varepsilon_{ij}({\bf E}) = \varepsilon_{ij}(0) + r_{ijk} E_k$ & \vsim V &  $RRR r^s$ \\
			(symmetric part) & $r^s_{ijk} = \frac{1}{2}(r_{ijk} + r_{jik})$ & & \\
			Ordinary Pockels effect & && \\
			\hline
			Pockels effect tensor $r^a_{ijk}$ & $\varepsilon_{ij}({\bf E}) = \varepsilon_{ij}(0) + r_{ijk} E_k$ & $a$\{V$^2$\}V &  $det(U)RRR r^a$ \\
			(antisymmetric part) & $r^a_{ijk} = \frac{1}{2}(r_{ijk} - r_{jik})$ &  &  \\
			\hline
		Faraday effect tensor $z^s_{ijk}$ & $\varepsilon_{ij}({\bf H}) = \varepsilon_{ij}(0) + i z_{ijk} H_k$ & $ae$\vsim V/\vsim M &  $RRU z^s$ (Spin)  \\
		(symmetric part) & $z^s_{ijk} = \frac{1}{2}(z_{ijk} + z_{jik})$ &  &  $\det(U)\det(R)RRR z^s$  \\
		Magnetooptic Kerr effect (MOKE) & && (Orbital)\\
		\hline
		Faraday effect tensor $z^a_{ijk}$ & $\varepsilon_{ij}({\bf H}) = \varepsilon_{ij}(0) + i z_{ijk} H_k$ & $e$\{V$^2$\}V/$a$\{V$^2$\}M &  $\det(U)RRU z^a$ (Spin) \\
		(antisymmetric part) & $z^a_{ijk} = \frac{1}{2}(z_{ijk} - z_{jik})$ &  & $\det(R)RRR z^a$  \\
		Ordinary Faraday effect &  &  & (Orbital) \\
		\hline
	\end{tabular}
\end{table}

 \clearpage
\subsection{Constraints imposed by collinearity and coplanarity on optical properties}
	\begin{table}[h]
	\centering
	\caption{Constraints imposed by collinearity and coplanarity on some tensors for optical properties.}
	\begin{tabular}{|l|l|l|}
		\hline
		\textbf{Tensor} & \textbf{Collinear structure} & \textbf{Coplanar structure} \\
		\hline
		Optical dielectric tensor $\varepsilon^s_{ij}$ & no restriction & no restriction \\
		(symmetric part) &  &  \\
		Index ellipsoid &  &  \\
		\hline
		Optical dielectric tensor $\varepsilon^a_{ij}$& $\varepsilon^a = 0$ & $\varepsilon^a = 0$ \\
		(antisymmetric part) &  &  \\
		Spontaneous Faraday effect &  &  \\
		\hline
		Optical activity tensor $\gamma^s_{ijk}$ & $\gamma^s = 0$ & $\gamma^s = 0$ \\
		(symmetric part) &  &  \\
		Spontaneous gyrotropic birefringence &  &  \\
		\hline
		Optical activity tensor $\gamma^a_{ijk}$ & no restriction & no restriction \\
		(antisymmetric part) &  &  \\
		Natural optical activity &  &  \\
		\hline
		Pockels effect tensor $r^s_{ijk}$  & no restriction & no restriction \\
		(symmetric part) &  &  \\
		Ordinary Pockels effect &  &  \\
		\hline
		Pockels effect tensor $r^a_{ijk}$ & $r^a = 0$ & $r^a = 0$ \\
		(antisymmetric part) &  &  \\
		\hline
		Faraday effect tensor $z^s_{ijk}$ & $z^s_{ij1} = z^s_{ij2} = 0$, & $z^s_{ij1}$, $z^s_{ij2}$ no restriction, \\
		(symmetric part, spin contribution) &  $z^s_{ij3}$ no restriction & $z^s_{ij3} = 0$ \\
		Magnetooptic Kerr effect (MOKE) &  &  \\
		\hline
		Faraday effect tensor $z^s_{ijk}$ & $z^s = 0$ & $z^s = 0$ \\
		(symmetric part, orbital contribution) & & \\
		Magnetooptic Kerr effect (MOKE) & & \\
		\hline
		Faraday effect tensor $z^a_{ijk}$ & $z^a = 0$ & $z^a_{ij1} = z^a_{ij2} = 0$, \\
		(antisymmetric part, spin contribution) &  &  $z^a_{ij3}$ no restriction \\
		Ordinary Faraday effect & & \\
		\hline
		Faraday effect tensor $z^a_{ijk}$ & no restriction & no restriction \\
		(antisymmetric part, orbital contribution) &  &  \\
		Ordinary Faraday effect &  &  \\
		\hline
	\end{tabular}
\end{table}
 \section{Nonlinear optical properties}
\label{s-nonlinear-properties}
Although there is a wide variety of nonlinear optical (NLO) properties, here we will study exclusively second-order electric-dipole processes which, when allowed, usually give the strongest signals. We will use the notation $\chi(\omega_3; \omega_2, \omega_1)$ to designate the NLO susceptibility in which input electric waves of frequencies $\omega_1$ and $\omega_2$ combine to produce an electric polarization at $\omega_3 = \omega_2 + \omega_1$, i.e.,
\begin{equation}
	\label{e-nonlinear-susceptibility}
	P_i(\omega_3) = \chi_{ijk}(\omega_3; \omega_2, \omega_1) E_j(\omega_2) E_k(\omega_1).
\end{equation}
This polarization, in its turn, produces an electric field with the same frequency $\omega_3$. Frequencies on the right of the semicolon (input waves) can be positive or negative; an input wave with negative frequency is equivalent to an output wave with a positive frequency. The frequency on the left side of the semicolon is the frequency of the output wave. It is always positive or zero.

As in the preceding cases, in order to obtain the restrictions produced by a SpPG we need the corresponding Onsager relations to find out the way these tensors behave under time reversal. In general, it turns out that the time reversal operation only gives a relation between elements of different NLO properties and is therefore not useful for tensor reduction \cite{Gallego2019}. Only in one special case Onsager relations can be exploited. This is the case $(\omega_2 = -\omega_1 = \omega, \omega_3 = 0)$, which corresponds to the so-called optical rectification phenomenon for which it can be shown \cite{Gallego2019} that
\begin{equation}
	\label{e-rectification}
	\{-1||1\}\chi_{ijk}(0;\omega,-\omega) = \chi_{ikj}(0;\omega,-\omega).
\end{equation}

From this expression we deduce that the symmetric part of this susceptibility in the last two indices $\left[\frac{1}{2}(\chi_{ijk}(0; \omega, -\omega) + \chi_{ikj}(0; \omega, -\omega))\right]$ is even with respect to $\{-1 || 1\}$, and the antisymmetric part $\left[\frac{1}{2}(\chi_{ijk}(0; \omega, -\omega) - \chi_{ikj}(0; \omega, -\omega))\right]$ is odd. This behavior together with the fact that the optical rectification tensor is polar implies that the symmetric part must be of type V\vsim~ and the antisymmetric part of type $a$V\{V$^2$\}.

As has been pointed out above, there are no more tensor symmetry reductions for the general case. However, further reductions can be attained in non-dissipative media because in those materials the NLO susceptibilities possess additional symmetries. More specifically, it can be shown that the absence of dissipation implies \cite{Pershan1963,Popov1995,Klyshko2011}
\begin{equation}
	\label{e-nondissipative}
	\{-1 || 1\} \chi_{ijk}(\omega_3; \omega_2, \omega_1) = \left[\chi_{ijk}(\omega_3; \omega_2, \omega_1)\right]^*,
\end{equation}
i.e., we retrieve a relation between elements of the same tensor property. Equation (\ref{e-nondissipative}) implies that the real part behaves like V\st~and the imaginary part is of type $a$V\st. The real part is interpreted physically as the contribution of the crystal lattice to $\chi$, while the imaginary part is understood as originating from the spin arrangement \cite{Gallego2019}. If in equation (\ref{e-nondissipative}) we take the special case of second harmonic generation, $(\omega_1 = \omega_2 = \omega, \omega_3 = 2\omega)$, the tensor is symmetric in its last two indices and we arrive at the symbols V\vsim~and $a$V\vsim~for the real and imaginary parts, respectively. If, additionally, the medium has no dispersion, $\chi_{ijk}$ has the so-called Kleinman symmetry \cite{Kleinman1962}, which allows any permutation of the indices $ijk$ in the real part (transforming then the symbol V\st~into [V$^3$]) and cancels out the imaginary part. Table S7 summarizes the situation in the different cases.

In this context, it is interesting to comment on a tensor with the same transformation properties as the SHG tensor in nondissipative media. This is the quadratic electrical conductivity tensor $\sigma_{ijk}$, which has recently been analyzed by \citet{Zhu2024}. Two main contributions can be distinguished in $\sigma_{ijk}$, the so-called Berry Curvature Dipole (BCD) contribution, which is V[V2], and the Quantum Metric Dipole (QMD) contribution, which is $a$V[V$^2$]. An additional part called the Inverse Mass Dipole (IMD) contribution also appears in the odd part for the time reversal, but it is of the type $a$[V$^3$], being then a special case of the QMD contribution.
It has been shown \cite{Tsirkin2022} that each contribution contains an Ohmic and a Hall-type (dissipationless) part, which can be separated according to the following prescription: The Ohmic parts correspond to the fully symmetric tensors [V$^3$] and $a$[V$^3$], while the Hall parts are those remaining after subtracting the Ohmic parts. The separation is carried out by imposing the following 10 constraints, $\sigma_{ijk} + \sigma_{jki} + \sigma_{kij} = 0$, on the tensors since, as can be easily checked, these conditions cancel out any tensor of type [V$^3$] or $a$[V$^3$].
 	\begin{table}
	\centering
	\caption{Summary of second-order electric-dipole susceptibilities with their Jahn symbols and transformation laws in the context of SpPGs.}
	\begin{tabular}{|l|l|l|l|}
		\hline
		\textbf{Tensor description} & \textbf{Range of validity} & \textbf{Jahn Symbol} & \textbf{Transformation} \\
		&  & \textbf{(MPG and SpPG)} & \textbf{law (SpPG)} \\
		\hline
		Optical rectification & General & V\vsim & $RRR\chi^s$ \\
		$\chi_{ijk}(0; \omega, -\omega)$ &  &  &  \\
		(symmetric part) &  &  &  \\
		\hline
		Optical rectification & General & $a$V\{V$^2$\} & $\det(U)RRR\chi^a$ \\
		$\chi_{ijk}(0; \omega, -\omega)$ &  &  &  \\
		(antisymmetric part) &  &  &  \\
		\hline
		General 2nd order susceptibility & Non dissipative media & V$^3$ & $RRR\chi^{real}$ \\
		$\chi_{ijk}(\omega_3; \omega_2, \omega_1)$ &  &  &  \\
		(real part) &  &  &   \\
		\hline
		General 2nd order susceptibility& Non dissipative media & $a$V$^3$ & $\det(U)RRR\chi^{imag}$ \\
		$\chi_{ijk}(\omega_3; \omega_2, \omega_1)$ &  &  &  \\
		(imaginary part) &  &  &  \\
		\hline
		Second-harmonic generation & Non dissipative media & V\vsim & $RRR\chi^{real}$ \\
		$\chi_{ijk}(2\omega; \omega, \omega)$&  &  & \\
		(real part) &  &  &  \\
		\hline
		Second-harmonic generation & Non dissipative media & $a$V\vsim & $\det(U)RRR\chi^{imag}$ \\
		$\chi_{ijk}(2\omega; \omega, \omega)$ &  &  &  \\
		(imaginary part) &  & & \\
		\hline
		General 2nd order susceptibility & Non dissipative and & [V$^3$] & $RRR\chi^{real}$ \\
		$\chi_{ijk}(\omega_3; \omega_2, \omega_1)$ & dispersionless media & & \\
		(real part) &  & & \\
		\hline
		General 2nd order susceptibility  & Non dissipative and & Forbidden & Forbidden \\
		$\chi_{ijk}(\omega_3; \omega_2, \omega_1)$ & dispersionless media & & \\
		(imaginary part) &  & & \\
		\hline
	\end{tabular}
\end{table} \section{Constraints on tensors for nonlinear optical susceptibilities of collinear and coplanar magnetic structures }
\label{s-nonlinear-constraints}
Collinearity and coplanarity also impose restrictions on the NLO susceptibility tensors. These restrictions are easily obtained because the Jahn symbols of all the properties for SpPG are not spin dependent and can be derived from the \mpgeff~(see Table S7). Especially remarkable are the vanishing of the antisymmetric part of the optical rectification tensor, and the imaginary parts of the second-order susceptibility tensor for non-dissipative media, both for collinear and coplanar structures. Table S8 summarizes these restrictions on some tensors for second-order NLO susceptibilities 	\begin{table}[ht]
	\centering
	\caption{Constraints imposed by collinearity and coplanarity on some tensors for second-order nonlinear optical properties.}
	\begin{tabular}{|l|l|l|}
		\hline
		\textbf{Tensor} & \textbf{Collinear structure} & \textbf{Coplanar structure} \\
		\hline
		Optical rectification $\chi_{ijk}(0; \omega, -\omega)$ & $\chi^s$ no restriction & $\chi^s$ no restriction \\
		(symmetric part) & & \\
		\hline
		Optical rectification $\chi_{ijk}(0; \omega, -\omega)$ & $\chi^a = 0$ & $\chi^a = 0$ \\
		(antisymmetric part) &  &  \\
		\hline
		General 2nd order susceptibility $\chi_{ijk}(\omega_3; \omega_2, \omega_1)$ & Re$\chi$ no restriction & Re$\chi$ no restriction \\
		(real part), (Non-dissipative media) &  &  \\
		\hline
		General 2nd order susceptibility $\chi_{ijk}(\omega_3; \omega_2, \omega_1)$ & Im$\chi = 0$ & Im$\chi = 0$ \\
		(imaginary part), (Non-dissipative media) & & \\
		\hline
		Second-harmonic generation $\chi_{ijk}(2\omega; \omega, \omega)$ & Re$\chi$ no restriction & Re$\chi$ no restriction \\
		(real part), (Non-dissipative media) & & \\
		\hline
		Second-harmonic generation $\chi_{ijk}(2\omega; \omega, \omega)$ & Im$\chi = 0$ & Im$\chi = 0$ \\
		(imaginary part), (Non-dissipative media) & & \\
		\hline
		General 2nd order susceptibility $\chi_{ijk}(\omega_3; \omega_2, \omega_1)$ & Re$\chi$ no restriction & Re$\chi$ no restriction \\
		(real part) & & \\
		(Non-dissipative and dispersionless media) &  &  \\
		\hline
		General 2nd order susceptibility $\chi_{ijk}(\omega_3; \omega_2, \omega_1)$ & Im$\chi = 0$ & Im$\chi = 0$ \\
		(imaginary part) & & \\
		(Non-dissipative and dispersionless media) &  &  \\
		\hline
	\end{tabular}
\end{table}
 \section{Study of further properties in non-coplanar DyVO$_3$ (entry 0.106 in MAGNDATA)}
	\label{s-DyVO3}	
	We finish the study of this material with a summary of results for two rank-3 tensors: the antisymmetric $R^a_{ijk}$(=$-R^a_{jik}$) and symmetric $R^s_{ijk}$(=$R^s_{jik}$) parts of the Hall tensor. The former is responsible for the ordinary Hall effect and the latter for the linear magnetoresistance. Analogous tensors (Table 6 in the main text and Table S5) also account for the ordinary Faraday effect and the magneto-optic Kerr effect respectively. 
	
	For the antisymmetric part (which is even with respect to time reversal) the reduction is as follows. There are 5 independent coefficients for the MPG ($R^a_{123}$, $R^a_{131}$, $R^a_{132}$, $R^a_{232}$, $R^a_{231}$). On the other hand, under the SpPG, there are 3 independent coefficients for the spin contribution ($R^a_{123}$, $R^a_{131}$, $R^a_{232}$), and 3 independent coefficients for the orbital part ($R^a_{123}$, $R^a_{231}$, $R^a_{132}$). Taking both contributions together we find no further reduction under the SpPG. Regarding the symmetric part $R^s_{ijk}$, which is odd with respect to the time reversal, the MPG allows 10 independent coefficients ($R^s_{111}$, $R^s_{121}$, $R^s_{221}$, $R^s_{331}$, $R^s_{112}$, $R^s_{122}$, $R^s_{222}$, $R^s_{332}$, $R^s_{133}$, $R^s_{233}$).  Under the SpPG only 5 of them survive for the spin contribution ($R^s_{121}$, $R^s_{112}$, $R^s_{222}$, $R^s_{332}$, $R^s_{133}$), and 5 for the orbital component ($R^s_{111}$, $R^s_{122}$, $R^s_{133}$, $R^s_{221}$, $R^s_{331}$). Taking both contributions together we only obtain one additional restriction ($R^s_{233}=0$) under the SpPG symmetry.
	
	In this example, it can be seen that some tensors of non-coplanar materials may not present many more constraints in the SpPGs than in the MPGs, especially if \Pso~is the trivial group.
 \section{Non-coplanar CaFe$_3$Ti$_4$O$_{12}$ (entry 3.24 in MAGNDATA)}
	\label{s-CaFe3Ti4O12}
\begin{figure}[ht]
	\centering
	\includegraphics[width=0.5\textwidth]{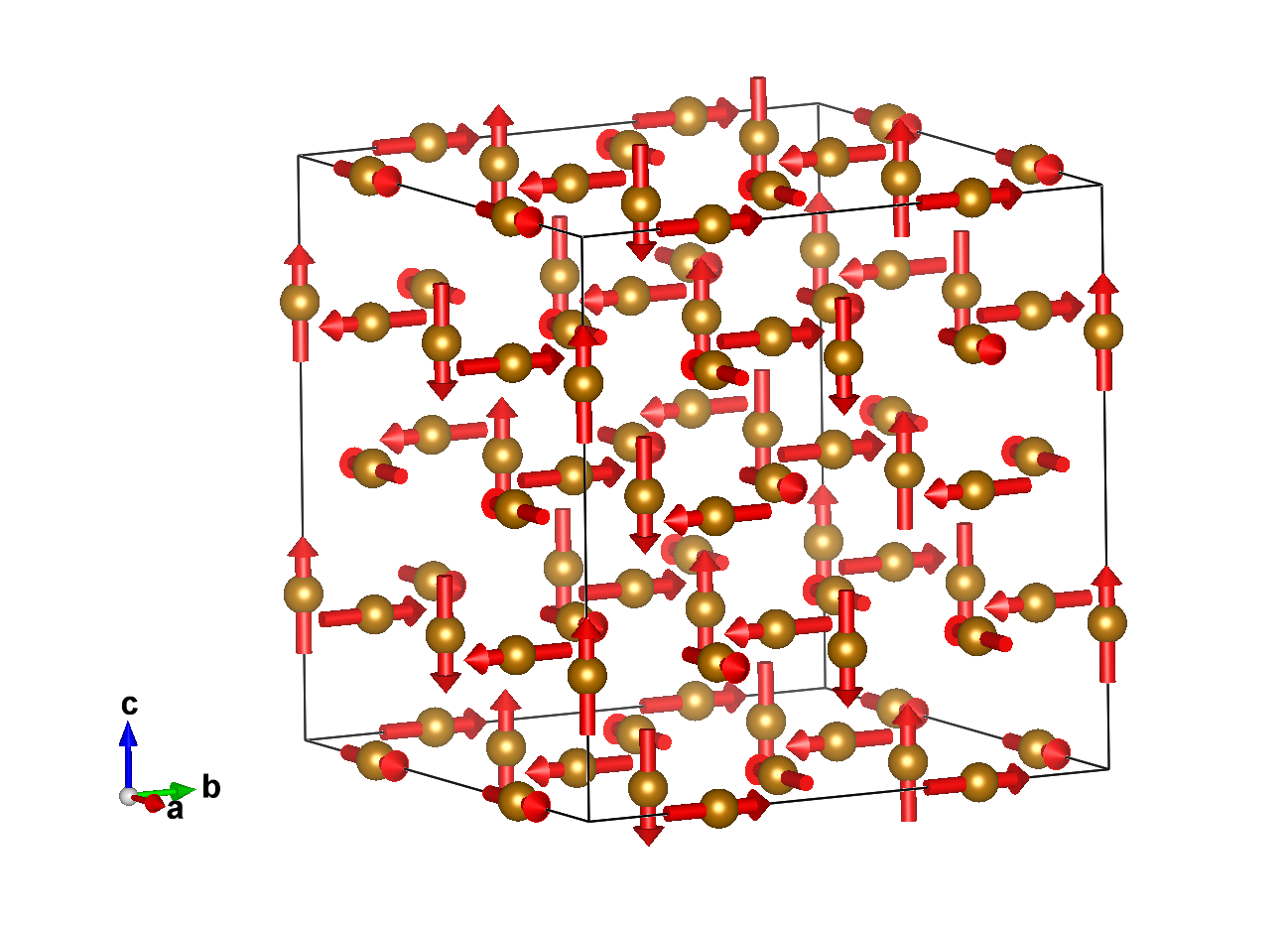}
	\caption{\label{f-CaFe3Ti4O12}Magnetic structure of CaFe$_3$Ti$_4$O$_{12}$ showing only the magnetic Fe and their spin orientation.}
\end{figure}

	The paramagnetic phase of CaFe$_3$Ti$_4$O$_{12}$ has space group $Im\bar{3}$ (No. 204) and the MSG of its magnetic phase is R$\bar{3}$ (OG N. 148.1.1247). The reported magnetic structure \cite{Patino2021} is shown in Fig. \ref{f-CaFe3Ti4O12}. Being a non-coplanar structure, the SpSG does not include any spin-only subgroup (except the identity), and coincides with its nontrivial subgroup, which is labelled with the numerical index 2.148.4.1 \cite{Chen2024}. Since $i_k = 4$, the SpSG cannot be minimal, despite the coincidence of the second number of the SpSG label and first number of the OG label of the MSG. In this case both G$_{\textrm{\footnotesize{0}}}$ and F are space groups of the same type ($R\overline{3}$), but the translation lattice of G$_{\textrm{\footnotesize{0}}}$ is denser than that of F. The corresponding SpPG is generated by the operations:
	\begin{displaymath}
		\{3_{111} ||\overline{3}_{111}\}, \{2_y || 1\},  \{2_z || 1\},
	\end{displaymath}
	while the MPG of the structure is $\bar{3}.1$, which has as single generator $\{3_{111} || \overline{3}_{111}\}$. It can be seen that the group of space operations is $\bar{3}$ in both cases, but the SpPG also includes a spin-only subgroup, \Pso, generated by $\{2_y || 1\}$ and $\{2_z || 1\}$. This spin-only group is originated by the spin-translation group present in the SpSG, and it can be denoted as $^{222}1$. Then the SpPG can be written as the direct product \Ps~= \Pm~$\times \,^{222}1$, with \Pm~being the MSG of the structure. We examine now the form of the magnetization, the anomalous Hall effect, and the spin Hall resistivity tensor allowed by the MPG and the SpPG.

	Magnetization is allowed under the MPG, with ${\bf M} = (M, M, M)$, i.e., along the trigonal axis. Under the SpPG, the spin contribution to the magnetization is forbidden due to the spin-only point group, but the \mpgeff~that dictates the constraints on the orbital contribution coincides with the MPG. Therefore, in the SOC-free approximation, any magnetization in the [111] direction can only be of orbital origin.
	
	On the other hand, the anomalous Hall effect (or the spontaneous Faraday effect) is permitted both by the MSG and SpPG, with the same form for the antisymmetric resistivity tensor:
	\begin{equation}
		\label{e-resistivity-example-3}
		\rho^a = \left(\begin{array}{ccc}
			0 & \rho_{12} & -\rho_{12} \\
			 -\rho_{12} & 0 & \rho_{12} \\
			\rho_{12} & -\rho_{12} & 0
		\end{array}\right)
	\end{equation}
	or the antisymmetric optical permittivity. Thus, in this material the anomalous Hall (and the Faraday) effect may be then of geometric nature.
	
	Finally, the MPG allows the existence of the spin Hall resistivity tensor, with 6 independent coefficients in its symmetric part:
	\begin{equation}
		\label{e-resistivity-example-3-symmetric}
		{\rho^s}^1=\left(\begin{array}{ccc}{\rho_{11}}^1&{\rho_{12}}^1&{\rho_{13}}^1\\{\rho_{12}}^1&-{\rho_{11}}^1&{\rho_{23}}^1\\{\rho_{13}}^1&{\rho_{23}}^1&0\end{array}\right),
		{\rho^s}^2=\left(\begin{array}{ccc}{\rho_{12}}^1&-{\rho_{11}}^1&-{\rho_{23}}^1\\-{\rho_{11}}^1&-{\rho_{12}}^1&{\rho_{13}}^1\\-{\rho_{23}}^1&{\rho_{13}}^1&0\end{array}\right),		
		{\rho^s}^3=\left(\begin{array}{ccc}{\rho_{11}}^3&0&0\\0&{\rho_{11}}^3&0\\0&0&{\rho_{33}}^3\end{array}\right)		
	\end{equation}
	and 3 coefficients in the antisymmetric part:
	\begin{equation}
	\label{e-resistivity-example-3-antisymmetric}
	{\rho^a}^1=\left(\begin{array}{ccc}0&0&{\rho_{13}}^1\\0&0&{\rho_{23}}^1\\-{\rho_{13}}^1&-{\rho_{23}}^1&0\end{array}\right),
	{\rho^a}^2=\left(\begin{array}{ccc}0&0&-{\rho_{23}}^1\\0&0&0\\{\rho_{23}}^1&0&0\end{array}\right),		
	{\rho^a}^3=\left(\begin{array}{ccc}0&{\rho_{12}}^3&0\\-{\rho_{12}}^3&0&0\\0&0&0\end{array}\right)		
\end{equation}
	However, according to the SpPG, all coefficients in equations (\ref{e-resistivity-example-3-symmetric}) and (\ref{e-resistivity-example-3-antisymmetric}) must be cancelled due to the constraints imposed by the spin-only group. Therefore, the whole property can only be an effect derived from the SOC. This major restriction deduced in the SpPG framework is due to the high symmetry of \Pso, and greatly contrasts with the preceding example.
 \section{Collinear UCr$_2$Si$_2$C (entry 0.499 in MAGNDATA)}
	
\begin{figure}[ht]
	\centering
	\includegraphics[width=0.5\textwidth]{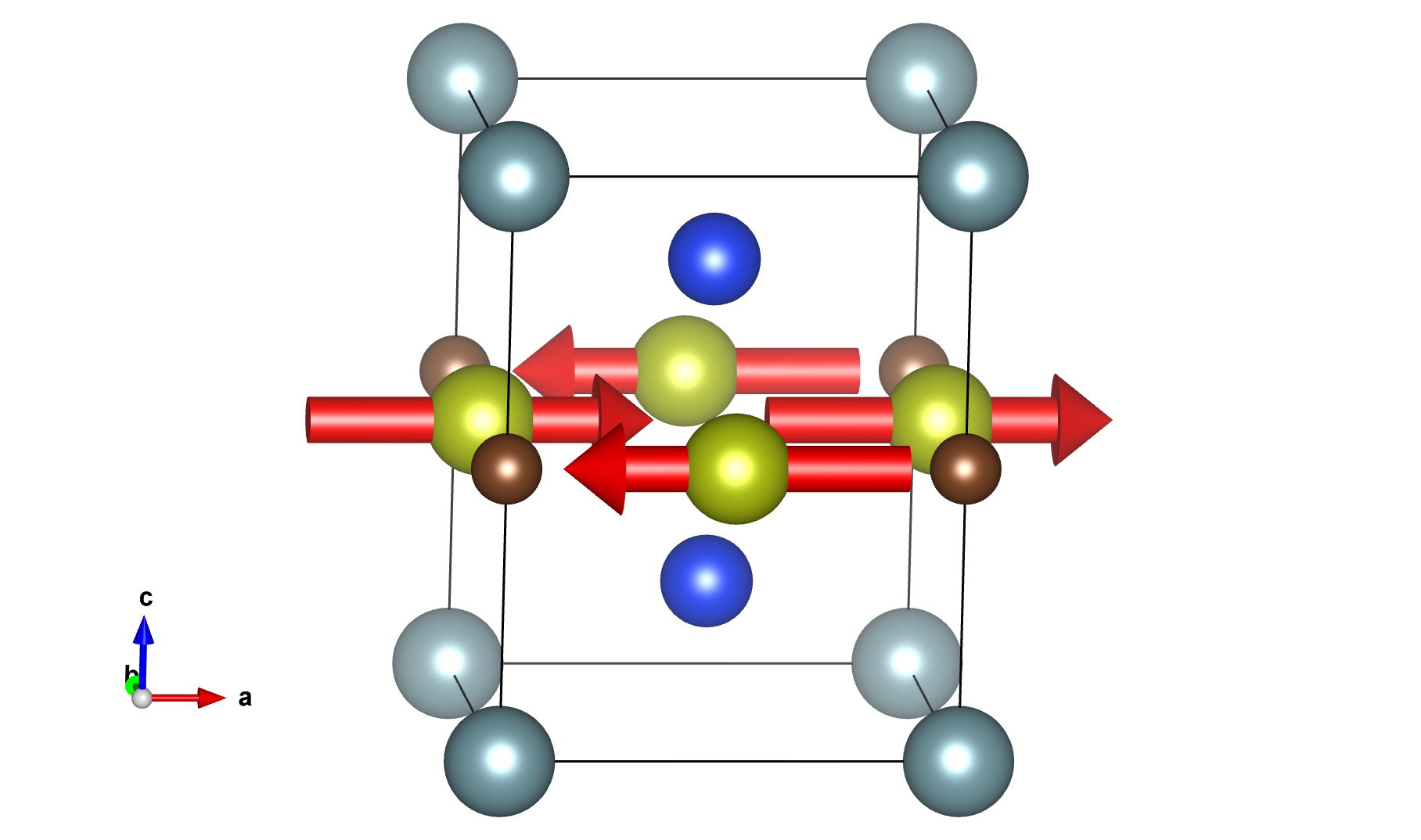}
	\caption{\label{f-UCr2Si2C}Magnetic structure of UCr$_2$Si$_2$C showing the spins of the Cr atoms (yellow spheres). The U, Si and C atoms are represented by gray, brown and blue colors respectively.}
\end{figure}

	UCr$_2$Si$_2$C has a tetragonal structure with space group $P4/mmm$ (No. 123) and it is magnetically ordered at room temperature. The reported collinear magnetic structure is shown in Fig. \ref{f-UCr2Si2C}. In the figure, the spin direction is taken along the $x$ axis of the tetragonal unit cell, but in fact it is only known that the spin direction is on the basal $xy$ plane, with its direction on the plane being experimentally undetermined \cite{Lemoine2018}. For the particular spin orientation assumed in Fig. \ref{f-UCr2Si2C}, and keeping the tetragonal crystallographic axes as the reference frame, the MSG of the structure can be written as $Pmm'm'$ (OG N. 47.4.350), and the corresponding MPG is $m_x m_y' m_z'$. On the other hand, the spin group symmetry, which is independent of the spin direction, is described by a collinear SpSG with nontrivial subgroup 47.123.1.1 \cite{Chen2024}. The corresponding SpPG is $^{-1}4/\,^{1}m\,^{1}m\,^{-1}m\,^{\infty m}1$. The SpPG and MPG are generated by the following elements (not a minimal set in the case of the SpPG to make more explicit the relation with the MPG generators):
	
	\begin{eqnarray}
		\textrm{SpPG} & : & \{1||m_z\}, \{1||m_y\}, \{1||\overline{1}\}, \{-1||4_z\}, \{\infty_x||1\}, \{m_z||1\}\nonumber\\
		\textrm{MPG}  & : & \{m_z||m_z\}, \{m_y||m_y\}, \{1||\overline{1}\}\nonumber
	\end{eqnarray}
 	
	Hence, while the SpPG is tetragonal with respect to the lattice operations, the MPG is only orthorhombic. For other spin orientations, say $x'$, the SpPG remains the same, with operations $\{\infty_{x'}||1\}$ instead of $\{\infty_x||1\}$, and $\{m_{z'}||1\}$ instead of $\{m_z||1\}$ ($z'$ is perpendicular to $x'$), whereas the MPG is reduced to $2'_z/m'_{z}$. An exception occurs if $x'$ is along $[110]$, in which case the MPG is $m_{1\bar{1}0} m'_{110} m'_{001}$.
	
	For the orbital contributions, the \mpgeff~is a gray group, $4/mmm.1'$, also independent of the spin direction.
	
	Tables \ref{t-UCr2Si2C-tensors} and \ref{t-UCr2Si2C-hall} show various tensor properties for 3 spin orientations within the easy plane ($[100]$, $[110]$, and $[210]$), and compare them with the tensors deduced under the MPG symmetry. When there are separate tensors with spin and orbital contributions both have been added. In general, one always finds higher reduction in the SpPG, in particular for the spin along $[210]$. An exception is the magnetic susceptibility, which shows both spin and orbital contributions. Especially remarkable are the cancellation of the spontaneous magnetization,  the antisymmetric part of the spin Hall resistivity, and the symmetric part of the spin Hall resistivity ${\rho^s}^3$ under the SpPG symmetry for all spin orientations. Note that as the SpPG is mathematically the same regardless of the spin direction, the number of resulting independent coefficients in each tensor is the same for all three spin orientations (Table \ref{t-UCr2Si2C-hall}), which is not the case when the MPG is considered.
	
\begin{table}[h]
	\centering
	\caption{\label{t-UCr2Si2C-tensors}Examples of symmetry-adapted tensors under MPG and SpPG symmetries for three different spin orientations in the collinear material UCr$_2$Si$_2$C.}
	\begin{tabular}{|c|c|c|c|}
		\hline
		\textbf{Spin} & \textbf{Tensor} & \textbf{MPG}         & \textbf{SpPG}         \\ 
		\textbf{direction} & \textbf{property} &         &         \\ 
		\hline
		100                     & Magnetization            & $(M_1, 0, 0)$        & $(0, 0, 0)$          \\ 
		\hline
		110                     & Magnetization            & $(M_1, -M_1, 0)$     & $(0, 0, 0)$          \\ 
		\hline
		210                     & Magnetization            & $(M_1, M_2, 0)$      & $(0, 0, 0)$          \\ 
		\hline
		100                     & Magnetic  &   \multirow{3}{*}{$\left(\begin{array}{ccc}\chi_{11}&0&0\\0&\chi_{22}&0\\0&0&\chi_{33}\end{array}\right)$}   &     \multirow{3}{*}{$\left(\begin{array}{ccc}\chi_{11}&0&0\\0&\chi_{22}&0\\0&0&\chi_{33}\end{array}\right)$}    \\ 
		                        & susceptibility  &                      &                      \\ 
		                        &   &                      &                      \\ 
		\hline
		110                     & Magnetic  &  \multirow{3}{*}{$\left(\begin{array}{ccc}\chi_{11}&\chi_{12}&0\\\chi_{12}&\chi_{11}&0\\0&0&\chi_{33}\end{array}\right)$}    &  \multirow{3}{*}{$\left(\begin{array}{ccc}\chi_{11}&\chi_{12}&0\\\chi_{12}&\chi_{11}&0\\0&0&\chi_{33}\end{array}\right)$} \\ 
		                     & susceptibility  &                      &                      \\ 
		                     &   &                      &                      \\ 
		\hline
		210                     & Magnetic  &  \multirow{3}{*}{$\left(\begin{array}{ccc}\chi_{11}&\chi_{12}&0\\\chi_{12}&\chi_{22}&0\\0&0&\chi_{33}\end{array}\right)$}  &   \multirow{3}{*}{$\left(\begin{array}{ccc}\chi_{11}&\chi_{12}&0\\\chi_{12}&\chi_{22}&0\\0&0&\chi_{33}\end{array}\right)$}  \\ 
		                     & susceptibility  &                      &                      \\ 
		                     &   &                      &                      \\ 
		\hline
		100                     & Electric     & \multirow{3}{*}{$\left(\begin{array}{ccc}\rho_{11}&0&0\\0&\rho_{22}&\rho_{23}\\0&-\rho_{23}&\rho_{33}\end{array}\right)$}  &  \multirow{3}{*}{$\left(\begin{array}{ccc}\rho_{11}&0&0\\0&\rho_{11}&0\\0&0&\rho_{33}\end{array}\right)$}    \\ 
		                     &  resistivity     &                      &                      \\ 
		                     &      &                      &                      \\ 
		\hline
		110                     & Electric     & \multirow{3}{*}{$\left(\begin{array}{ccc}\rho_{11}&\rho_{12}&\rho_{13}\\\rho_{12}&\rho_{11}&\rho_{23}\\-\rho_{13}&-\rho_{23}&\rho_{33}\end{array}\right)$}  &  \multirow{3}{*}{$\left(\begin{array}{ccc}\rho_{11}&0&0\\0&\rho_{11}&0\\0&0&\rho_{33}\end{array}\right)$}    \\ 
210&  resistivity     &                      &                      \\ 
&      &                      &                      \\ 
		\hline
	\end{tabular}
\end{table}
 	
	\begin{table}[h]
	\centering
	\caption{\label{t-UCr2Si2C-hall}Symmetry-adapted tensors under MPG and SpPG symmetries of the symmetric and antisymmetric parts of the spin Hall resistivity (SHR) tensor ${\rho_{ij}}^k$ for 3 different spin orientations in the collinear material UCr$_2$Si$_2$C.}
	\begin{footnotesize}
		\begin{tabular}{|c|c|c|}
			\hline
			{\bf SHR} & {\bf MPG} & {\bf SpPG} \\
			{\bf tensor} &  &  \\
			\hline
			Symmetric & \multirow{3}{*}{$\left(\begin{array}{ccc}{\rho_{11}}^1&0&0\\0&{\rho_{22}}^1&0\\0&0&{\rho_{33}}^1\end{array}\right),
				\left(\begin{array}{ccc}0&{\rho_{12}}^2&0\\{\rho_{12}}^2&0&0\\0&0&0\end{array}\right),$} & \multirow{3}{*}{$\left(\begin{array}{ccc}{\rho_{11}}^1&0&0\\0&-{\rho_{11}}^1&0\\0&0&0\end{array}\right),$}\\
			part [100] & & \\
			& & \\
			& \multirow{3}{*}{$\left(\begin{array}{ccc}0&0&{\rho_{13}}^3\\0&0&0\\{\rho_{13}}^3&0&0\end{array}\right)$}& $\rho^2=\rho^3=0$\\
			& & \\
			& & \\
			\hline
			Antiymmetric & \multirow{3}{*}{$\left(\begin{array}{ccc}0&0&0\\0&0&{\rho_{23}}^1\\0&-{\rho_{23}}^1&0\end{array}\right),
				\left(\begin{array}{ccc}0&0&{\rho_{13}}^2\\0&0&0\\-{\rho_{13}}^2&0&0\end{array}\right),$} & \multirow{3}{*}{$\rho^1=\rho^2=\rho^3=0$}\\
			part [100] & & \\
			& & \\
			& \multirow{3}{*}{$\left(\begin{array}{ccc}0&{\rho_{12}}^3&0\\-{\rho_{12}}^3&0&0\\0&0&0\end{array}\right)$}& \\
			& & \\
			& & \\
			\hline
			Symmetric & \multirow{3}{*}{$\left(\begin{array}{ccc}{\rho_{11}}^1&{\rho_{12}}^1&0\\{\rho_{12}}^1&{\rho_{22}}^1&0\\0&0&{\rho_{33}}^1\end{array}\right),
				\left(\begin{array}{ccc}-{\rho_{22}}^1&-{\rho_{12}}^1&0\\-{\rho_{12}}^1&-{\rho_{11}}^1&0\\0&0&-{\rho_{33}}^1\end{array}\right),$} & \multirow{3}{*}{$\left(\begin{array}{ccc}{\rho_{11}}^1&0&0\\0&-{\rho_{11}}^1&0\\0&0&0\end{array}\right),\left(\begin{array}{ccc}{\rho_{11}}^1&0&0\\0&-{\rho_{11}}^1&0\\0&0&0\end{array}\right),$}\\
			part [110] & & \\
			& & \\
			& \multirow{3}{*}{$\left(\begin{array}{ccc}0&0&{\rho_{13}}^3\\0&0&-{\rho_{13}}^3\\{\rho_{13}}^3&-{\rho_{13}}^3&0\end{array}\right)$}& $\rho^3=0$\\
			& & \\
			& & \\
			\hline
			Antiymmetric & \multirow{3}{*}{$\left(\begin{array}{ccc}0&0&{\rho_{13}}^1\\0&0&{\rho_{23}}^1\\-{\rho_{13}}^1&-{\rho_{23}}^1&0\end{array}\right),
				\left(\begin{array}{ccc}0&0&-{\rho_{23}}^1\\0&0&-{\rho_{13}}^1\\{\rho_{23}}^1&{\rho_{13}}^1&0\end{array}\right),$} & $\rho^1=\rho^2=\rho^3=0$\\
			part [110] & & \\
			& & \\
			& \multirow{3}{*}{$\left(\begin{array}{ccc}0&{\rho_{12}}^3&0\\-{\rho_{12}}^3&0&0\\0&0&0\end{array}\right)$}& \\
			& & \\
			& & \\
			\hline
			Symmetric & \multirow{3}{*}{$\left(\begin{array}{ccc}{\rho_{11}}^1&{\rho_{12}}^1&0\\{\rho_{12}}^1&{\rho_{22}}^1&0\\0&0&{\rho_{33}}^1\end{array}\right),
				\left(\begin{array}{ccc}{\rho_{11}}^2&{\rho_{12}}^2&0\\{\rho_{12}}^2&{\rho_{22}}^2&0\\0&0&{\rho_{33}}^2\end{array}\right),$} & \multirow{3}{*}{$\left(\begin{array}{ccc}{\rho_{11}}^1&0&0\\0&-{\rho_{11}}^1&0\\0&0&0\end{array}\right),\left(\begin{array}{ccc}\frac{{\rho_{11}}^1}{2}&0&0\\0&-\frac{{\rho_{11}}^1}{2}&0\\0&0&0\end{array}\right),$}\\
			part [210] & & \\
			& & \\
			& \multirow{3}{*}{$\left(\begin{array}{ccc}0&0&{\rho_{13}}^3\\0&0&{\rho_{23}}^3\\{\rho_{13}}^3&{\rho_{23}}^3&0\end{array}\right)$}& $\rho^3=0$\\
			& & \\
			& & \\
			\hline
			Antiymmetric & \multirow{3}{*}{$\left(\begin{array}{ccc}0&0&{\rho_{13}}^1\\0&0&{\rho_{23}}^1\\-{\rho_{13}}^1&-{\rho_{23}}^1&0\end{array}\right),
				\left(\begin{array}{ccc}0&0&{\rho_{13}}^2\\0&0&{\rho_{23}}^2\\-{\rho_{13}}^2&-{\rho_{23}}^2&0\end{array}\right),$} & $\rho^1=\rho^2=\rho^3=0$\\
			part [210] & & \\
			& & \\
			& \multirow{3}{*}{$\left(\begin{array}{ccc}0&{\rho_{12}}^3&0\\-{\rho_{12}}^3&0&0\\0&0&0\end{array}\right)$}& \\
			& & \\
			& & \\
			\hline
		\end{tabular}
	\end{footnotesize}
\end{table}

\clearpage
\bibliography{references_arXiv}
\end{document}